\documentclass[trackchanges,twocolumn]{aastex701}

\usepackage{amsmath}

\begin{document}

\correspondingauthor{Shuo Zhai}
\email{zhaishuo@bao.ac.cn}

\title{A Decade-Scale Ionization Echo of Black-Hole Accretion in Galactic Nuclei}

\author[0000-0001-9457-0589]{Wei-Jian Guo}
\affiliation{National Astronomical Observatories, Chinese Academy of Sciences,
20A Datun Road, Chaoyang District, Beijing 100101,
People's Republic of China}
\affiliation{Centre for Extragalactic Astronomy, Department of Physics,
Durham University, South Road, Durham DH1 3LE, UK}
\email{guowj@bao.ac.cn}

\author[0009-0005-4152-2088]{Shuo Zhai}
\affiliation{National Astronomical Observatories, Chinese Academy of Sciences,
20A Datun Road, Chaoyang District, Beijing 100101,
People's Republic of China}
\email{zhaishuo@bao.ac.cn}

\author{Chen Hu}
\affiliation{Key Laboratory for Particle Astrophysics,
Institute of High Energy Physics, Chinese Academy of Sciences,
19B Yuquan Road, Beijing 100049,
People's Republic of China}
\email{huc@ihep.ac.cn}

\author[0000-0002-5896-6313]{David M. Alexander}
\affiliation{Centre for Extragalactic Astronomy, Department of Physics,
Durham University, South Road, Durham DH1 3LE, UK}
\affiliation{Institute for Computational Cosmology, Department of Physics,
Durham University, South Road, Durham DH1 3LE, UK}
\email{d.m.alexander@durham.ac.uk}

\author[0000-0002-9036-0063]{Bin Luo}
\affiliation{School of Astronomy and Space Science,
Nanjing University, Nanjing 210023,
People's Republic of China}
\email{bluo@nju.edu.cn}

\author[0000-0002-5830-3544]{Pu Du}
\affiliation{Key Laboratory for Particle Astrophysics,
Institute of High Energy Physics, Chinese Academy of Sciences,
19B Yuquan Road, Beijing 100049,
People's Republic of China}
\email{dupu@ihep.ac.cn}

\author[0000-0001-8433-550X]{Matthew J. Temple}
\affiliation{Centre for Extragalactic Astronomy, Department of Physics,
Durham University, South Road, Durham DH1 3LE, UK}
\email{matthew.j.temple@durham.ac.uk}

\author[0000-0003-0487-6651]{Leah K. Morabito}
\affiliation{Centre for Extragalactic Astronomy, Department of Physics,
Durham University, South Road, Durham DH1 3LE, UK}
\affiliation{Institute for Computational Cosmology, Department of Physics,
Durham University, South Road, Durham DH1 3LE, UK}
\email{leah.k.morabito@durham.ac.uk}

\author[0000-0002-7633-431X]{Feige Wang}
\affiliation{Department of Astronomy, University of Michigan,
1085 S. University Avenue, Ann Arbor, MI 48109-1107, USA}
\email{fgwang@umich.edu}

\author[0000-0001-5287-4242]{Jinyi Yang}
\affiliation{Department of Astronomy, University of Michigan,
1085 S. University Avenue, Ann Arbor, MI 48109-1107, USA}
\email{jyyangas@umich.edu}

\author[0000-0003-1251-532X]{Victoria A. Fawcett}
\affiliation{European Southern Observatory,
Karl-Schwarzschild-Straße 2,
85748 Garching bei München, Germany}
\email{Victoria.Fawcett@eso.org}

\author[0000-0002-5854-7426]{Swayamtrupta Panda}
\affiliation{International Gemini Observatory/NSF NOIRLab,
Casilla 603, La Serena, Chile}
\email{swayamtrupta.panda@noirlab.edu}

\author[0000-0002-7719-5809]{Claire L. Greenwell}
\affiliation{Centre for Extragalactic Astronomy, Department of Physics,
Durham University, South Road, Durham DH1 3LE, UK}
\email{claire.l.greenwell@durham.ac.uk}

\author[0000-0001-5841-9179]{Yan-Rong Li}
\affiliation{Key Laboratory for Particle Astrophysics,
Institute of High Energy Physics, Chinese Academy of Sciences,
19B Yuquan Road, Beijing 100049,
People's Republic of China}
\email{liyanrong@ihep.ac.cn}

\author[0000-0003-0202-0534]{Cheng Cheng}
\affiliation{National Astronomical Observatories, Chinese Academy of Sciences,
20A Datun Road, Chaoyang District, Beijing 100101,
People's Republic of China}
\email{chengcheng@nao.cas.cn}

\author[0000-0002-1234-552X]{Shengxiu Sun}
\affiliation{Kavli Institute for Astronomy and Astrophysics
at Peking University,
Peking University, 5 Yiheyuan Road, Haidian District,
Beijing 100871, People's Republic of China}
\email{sxsun@stu.pku.edu.cn}

\author[0009-0008-3338-393X]{Zhi-Qiang Chen}
\affiliation{School of Physics and Technology,
Nanjing Normal University,
No. 1 Wenyuan Road, Nanjing 210023,
People's Republic of China}
\email{chenzhiqiang@nnu.edu.cn}

\author[0000-0002-9889-4238]{Alessandro Marconi}
\affiliation{{Dipartimento di Fisica e Astronomia,
Universit\`a di Firenze, Via G. Sansone 1,
50019 Sesto F.no (Firenze), Italy}}
\affiliation{{INAF-Osservatorio Astrofisico di Arcetri,
Largo E. Fermi 5, 50125 Firenze, Italy}}
\email{}

\author[0000-0001-9449-9268]{Jian-Min Wang}
\affiliation{Key Laboratory for Particle Astrophysics,
Institute of High Energy Physics, Chinese Academy of Sciences,
19B Yuquan Road, Beijing 100049,
People's Republic of China}
\affiliation{National Astronomical Observatories, Chinese Academy of Sciences,
20A Datun Road, Chaoyang District, Beijing 100101,
People's Republic of China}
\affiliation{School of Astronomy and Space Science,
University of Chinese Academy of Sciences,
Beijing 100049, People's Republic of China}
\email{wangjm@ihep.ac.cn}

\begin{abstract}
Circumnuclear structures in active galactic nuclei respond to accretion variability over different spatial and temporal scales, but whether compact narrow-line gas responds on decade timescales has not been systematically established. We use repeat Sloan Digital Sky Survey and Dark Energy Spectroscopic Instrument spectroscopy of 473 variability-selected galactic nuclei with evidence for recent nuclear brightening. We identify 19 sources whose narrow [O~III] $\lambda5007$ flux increased by more than a factor of two over rest-frame baselines of 12.0--20.4 yr, which we term narrow-line ionization echoes (NIEs). In equal-weight stacked spectra, [O~III] increased by $229\pm22\%$, while lower-ionization lines remained nearly unchanged or declined, indicating an ionization-dependent response rather than uniform spectral rescaling. The NIE hosts also moved coherently toward higher excitation in optical diagnostic diagrams, with mean increases of $0.46\pm0.18$ dex in $\log([\mathrm{O~III}]/\mathrm{H}\beta)$ and $0.58\pm0.13$ dex in $\log([\mathrm{O~III}]/[\mathrm{O~II}])$. Photoionization models reproduce this evolution with a higher ionization parameter and a harder AGN-like ionizing spectrum at fixed gas abundance. For 13 sources with well-defined mid-infrared outbursts, the integrated W1+W2 excess energy correlates strongly with the absolute [O~III] luminosity change ($r=0.87$), linking the dust and narrow-line responses to the same nuclear event. The mid-infrared timescale and spectroscopic baseline constrain the variable [O~III]-emitting gas to characteristic scales of 1.5--5.2 pc. Compact narrow-line gas therefore appears to form an intermediate, decade-scale response layer between the infrared dust echo and the long-lived extended narrow-line region.
\end{abstract}


\section{Introduction}\label{sec:intro}

Accretion onto massive black holes is intrinsically variable.
Changes in the accretion flow can produce large variations in the
ionizing continuum of active galactic nuclei (AGNs), ranging from
variability within a persistent accretion flow to major accretion-state
changes and transient fueling events. The latter may include the tidal
disruption of stars or nuclear transients whose physical origins remain
uncertain
\citep[e.g.,][]{Ulrich1997,Rees1988,Gezari2021}.
Radiation from such an episode is absorbed and reprocessed by the gas
and dust surrounding the nucleus. Owing to finite light-travel and
recombination times, the reprocessed emission can persist after the
direct nuclear radiation has declined. These delayed signals therefore
provide information on recent black-hole accretion activity even when
the initiating event was short-lived, obscured, or incompletely
observed.

The response of circumnuclear material is stratified in both radius and
timescale. Broad-line-emitting gas responds to continuum variations on
timescales of days to months, forming the basis of reverberation mapping
of the innermost line-emitting region
\citep[e.g.,][]{Peterson1993,Cackett2021}.
Circumnuclear dust absorbs ultraviolet and optical radiation and
reradiates it in the mid-infrared (MIR), producing echoes on timescales
of months to years
\citep[e.g.,][]{Suganuma2006,Koshida2014,Lyu2019}.
The narrow-line region (NLR) extends from parsec to kiloparsec scales
and contains gas spanning a broad range of densities and ionization
states
\citep[e.g.,][]{Bennert2002,Schmitt2003,Bennert2006,Hainline2013,
Liu2013}.
At the largest scales, spatially resolved ionization echoes, including
Hanny's Voorwerp and other extended AGN-ionized clouds, require an
ionizing luminosity substantially greater than that of the currently
observed nucleus
\citep{Lintott2009,Keel2012,Schirmer2016,Keel2017}.
These systems preserve records of nuclear activity over
$\sim10^{4}$--$10^{5}$ yr
\citep{Schweizer2013,Schawinski2015}.
The response of compact, spatially unresolved narrow-line gas on
intermediate timescales of years to decades, however, remains poorly
characterized at the population level. An unresolved narrow-line
spectrum may consequently contain emission from gas responding to a
recent change in the ionizing continuum together with emission from
more extended gas that reflects the longer-term nuclear output.

Evidence for variability from compact narrow-line gas has been reported
in several contexts. Long-term monitoring of NGC~5548 revealed changes
in its narrow [\ion{O}{3}] emission and constrained the variable
component to parsec scales
\citep{Peterson2013}.
Extreme coronal-line emitters, many of which have been associated with
tidal disruption events or related nuclear flares, show substantial
spectral evolution over several years. Their evolution includes the
fading of high-ionization coronal lines and, in some objects, the later
strengthening of lower-ionization coronal lines and [\ion{O}{3}]
\citep{Wang2012,Yang2013,Clark2024,Short2023,
Newsome2024,Hinkle2024,Kynoch2026}.
Large changes in an existing accretion flow, including turn-on and
changing-look transitions, can also modify the ionizing radiation
incident on circumnuclear gas
\citep{Guo2024CLAGN,Guo2025CLAGN}.
These results show that forbidden-line emission can respond to nuclear
activity on observable timescales and that similar responses may follow
different classes of nuclear events. Most known examples, however,
have been identified through long-term monitoring of individual AGNs,
selection based on unusually strong coronal lines, or imaging of
extended emission-line nebulae. It remains unclear whether comparable
responses can be identified systematically using the strong optical
narrow lines available in large spectroscopic surveys.

The response of the optical narrow-line spectrum is expected to depend
on ionization state. The different forbidden lines arise from gas
occupying distinct ionization zones and need not respond by the same
amount to a change in the incident radiation field. In particular,
[\ion{O}{3}] $\lambda5007$ traces O$^{2+}$ and is sensitive to the
supply of photons above the O$^{+}$ ionization threshold, whereas
[\ion{O}{2}], [\ion{N}{2}], and [\ion{S}{2}] preferentially trace
lower-ionization gas
\citep{Osterbrock2006}.
An increase in the ionizing flux or a hardening of the ionizing
spectrum can therefore enhance [\ion{O}{3}] relative to the
lower-ionization lines. The corresponding changes in line ratios should
appear as systematic evolution toward higher excitation in standard
optical diagnostic diagrams
\citep{BPT1981,Kewley2001,
Kauffmann2003,Kewley2006}.
Repeat spectroscopy can thus reveal temporal changes in the ionization
structure of unresolved narrow-line-emitting gas.

MIR variability provides an independent tracer of the nuclear
brightening that precedes the narrow-line response. Circumnuclear dust
reradiates absorbed ultraviolet and optical emission in the infrared,
allowing a nuclear event to remain visible when the direct optical
continuum is obscured, diluted by host-galaxy starlight, or has already
faded
\citep{Sheng2017,Lyu2019,vanVelzen2021}.
MIR light curves and repeat optical spectroscopy therefore trace the
responses of two distinct components of the circumnuclear environment.
The combination of WISE/NEOWISE time-domain photometry with
spectroscopy from the Sloan Digital Sky Survey and the Dark
Energy Spectroscopic Instrument  now permits these responses to
be studied over large samples and decade-long baselines
\citep{York2000,DESI2024,Wright2010,Mainzer2011,
Mainzer2014}.

In this work, we combine WISE/NEOWISE light curves with repeat SDSS and
DESI spectroscopy to search for delayed narrow-line responses to recent
nuclear activity. We construct a parent sample of 473 galactic nuclei
at $z<0.35$ from published catalogs of MIR-variable galaxies, turn-on
AGNs, and dust-echo-like infrared flares
\citep{Jiang2021,Yang2025,Necker2025}.
The parent sample is selected on the basis of recent nuclear activity,
independently of the optical narrow-line changes examined here.
Applying a uniform decomposition to the SDSS and DESI spectra, we
identify 19 sources in which the narrow [\ion{O}{3}] $\lambda5007$
flux increases by more than a factor of two. We refer to these objects
as narrow-line ionization echo (NIE) hosts. The term denotes the delayed
response of narrow-line-emitting gas and does not assign a specific
physical origin to the initiating ionizing event. The NIE hosts exhibit
an ionization-ordered narrow-line response rather than a uniform
rescaling of the narrow-line spectrum.

The remainder of this paper is organized as follows.
Section~\ref{sec:data} describes the spectroscopic and time-domain
datasets used in this work. Section~\ref{sec:nie_identification} presents the
construction of the variability-selected parent sample, the
identification of NIE hosts, and the validation of their narrow-line
variability. Section~\ref{sec:ionization_evolution} investigates the
ionization-state evolution of the narrow-line gas and presents the
photoionization modeling. Section~\ref{sec:infrared} describes the
mid-infrared variability, characteristic timescales, and connection
between the dust-reprocessed emission and the narrow-line response.
Section~\ref{sec:discussion} discusses the physical scale and
structure of the responsive gas, the possible origins of the ionizing
episodes, and the broader ionization-memory framework. We summarize
our main conclusions in Section~\ref{sec:conclusions}. Throughout this work, we adopt a flat $\Lambda$CDM cosmology with
$H_{0}=67.4\ \mathrm{km\,s^{-1}\,Mpc^{-1}}$,
$\Omega_{\rm m}=0.315$, and $\Omega_{\Lambda}=0.685$
\citep{Planck2020}.

\begin{figure*}[t]
    \centering
    \includegraphics[width=0.8\textwidth]
    {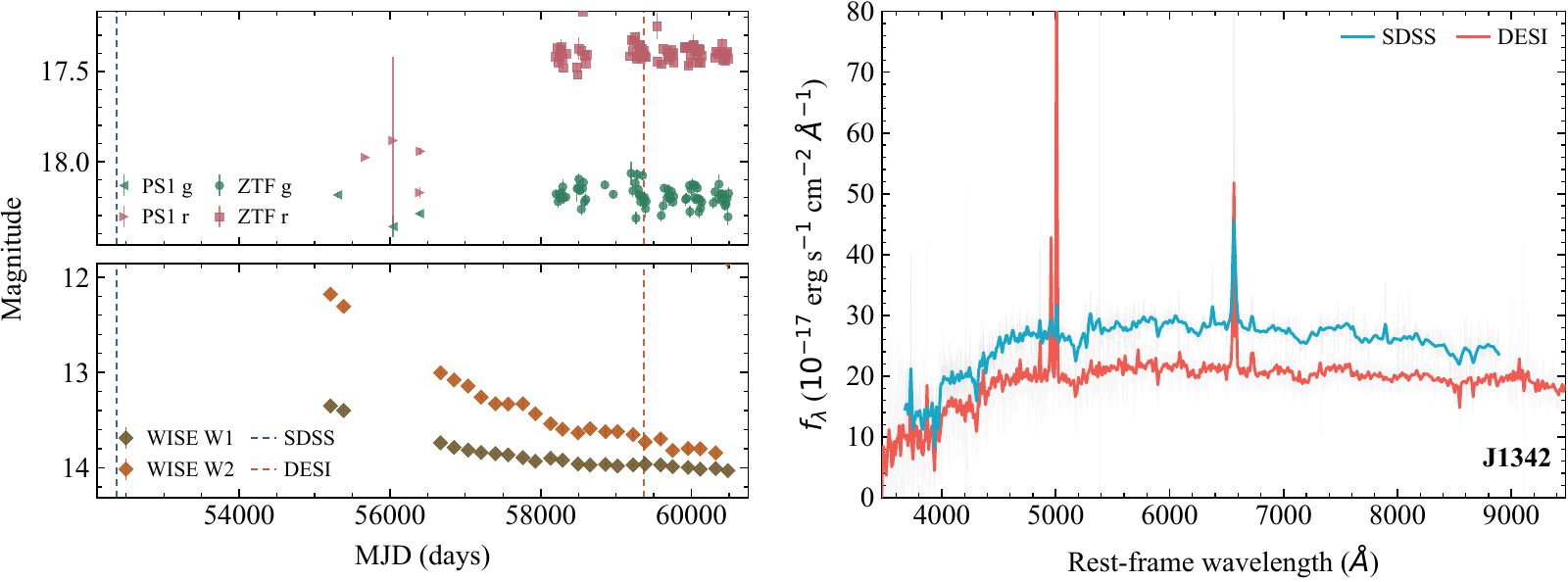}
    \caption{\textbf{Multiwavelength variability and optical spectral evolution of J1342.}
    The left panels show the optical (top) and mid-infrared (bottom) light curves of J134244.42+053056.1. Green and muted-red points denote the $g$- and $r$-band measurements from Pan-STARRS1 and ZTF, while brown and orange symbols represent the WISE W1 and W2 measurements, respectively. The blue and red dashed vertical lines mark the SDSS and DESI spectroscopic epochs. The right panel compares the SDSS (blue) and DESI (red) optical spectra in the rest frame. Both spectra were corrected for Galactic extinction using $E(B-V)=0.0249$ and $R_V=3.1$, and transformed to the rest frame adopting $z=0.0366$. The faint gray curves show the corrected pixel-level spectra, while the colored curves show Gaussian-smoothed versions for visualization. No relative flux rescaling was applied between the two epochs. The later DESI spectrum exhibits substantially enhanced narrow [O~III] emission despite a lower optical continuum, while the long-term WISE light curves place this spectral transition in the context of pronounced mid-infrared variability.}
    \label{fig:j1342}
\end{figure*}

\section{Spectroscopy and Light Curve}
\label{sec:data}

\subsection{SDSS and DESI Spectroscopy}
\label{subsec:spectroscopic_data}
The earlier-epoch spectra were obtained from the Sloan Digital Sky
Survey \citep[SDSS;][]{York2000,Ahumada2020}. SDSS is a long-running
imaging and spectroscopic survey conducted with the 2.5-m Sloan
Foundation Telescope at Apache Point Observatory and has established
one of the principal optical spectroscopic archives for low-redshift
galaxies and quasars. The spectra used here were obtained through
$3\arcsec$-diameter or  $2\arcsec$-diameter fibers, with an observed-frame wavelength coverage of approximately $3800$--$9200$~\AA\ and a resolving power of
$R\sim2000$.

The later-epoch spectra were obtained with the Dark Energy
Spectroscopic Instrument
\citep[DESI;][]{DESI2016,DESI_2022,DESI_DR1}. DESI is a highly
multiplexed spectroscopic survey instrument installed on the 4-m
Nicholas U. Mayall Telescope at Kitt Peak National Observatory. Its
5000 robotically positioned fibers ($1.5\arcsec$-diameter) cover a field of approximately
$8~\mathrm{deg}^{2}$ in each pointing and feed ten three-arm
spectrographs, providing continuous wavelength coverage from
approximately $3600$ to $9800$~\AA\ at a resolving power of
$R\sim2000$--$5000$. Although designed primarily to map large-scale
structure, its broad target selection and overlap with SDSS provide a
large sample of decade-separated repeat spectra for studies of galaxy
and nuclear variability.

The SDSS and DESI spectra differ in fiber diameter,
spectral resolution, and spectrophotometric calibration; these effects
are assessed using the full parent population in
Section~\ref{sec:nie_identification}. Before spectral modeling, all spectra
were corrected for Galactic extinction using the
\citet{Schlegel1998} dust map as implemented in
\texttt{dustmaps} \citep{Green2018}, adopting the
\citet{Fitzpatrick1999} extinction curve with $R_V=3.1$. DESI
redshifts were used to shift both epochs to the rest frame and to
calculate luminosity distances.

\subsection{WISE/NEOWISE Photometry}
\label{subsec:wise_data}

We used MIR photometry from the
\textit{Wide-field Infrared Survey Explorer}
(\textit{WISE}) and its NEOWISE reactivation mission
\citep{Wright2010,Mainzer2011,Mainzer2014}.
The W1 and W2 bands, centered at approximately 3.4 and
$4.6~\micron$, respectively, trace nuclear radiation absorbed and
reradiated by circumnuclear dust. Their long temporal baseline
therefore provides a record of recent nuclear activity that is
independent of the optical narrow-line changes measured from the SDSS
and DESI spectra.

For each source, we retrieved the single-exposure W1 and W2
measurements and grouped observations obtained within the same
observing period into epochs. The resulting epoch-binned light curves
were used to characterize the long-term MIR variability of the parent
sample and the NIE hosts. Where the rise and decline of an MIR
outburst were sufficiently constrained, the light curves were also
used to define the outburst interval required for the timescale and
excess-energy measurements presented in
Section~\ref{sec:infrared}. The photometric quality selection, epoch
binning, baseline determination, and conversion from magnitude to
luminosity are described there.

\subsection{Optical Photometric Data}
\label{subsec:optical_photometry}
We additionally compiled optical photometry from the Zwicky Transient
Facility
\citep[ZTF;][]{Bellm2019,Masci2019}
and Pan-STARRS
\citep{Chambers2016}.
These data were used to place the SDSS and DESI spectroscopic epochs
within the broader optical variability history of each source and to
examine whether the MIR activity was accompanied by a detectable
change in the direct optical emission.

The optical light curves provide contextual information only. They
were not used to select the NIE hosts, measure the narrow emission
lines, define the MIR integration intervals, or calculate the MIR
excess energies. This separation is particularly relevant for
host-dominated or obscured nuclei, in which a substantial nuclear
outburst may produce a prominent MIR dust echo while the corresponding
optical continuum variation remains weak or undetectable.

\begin{figure*}[t]
\centering
\includegraphics[width=0.9\linewidth]
{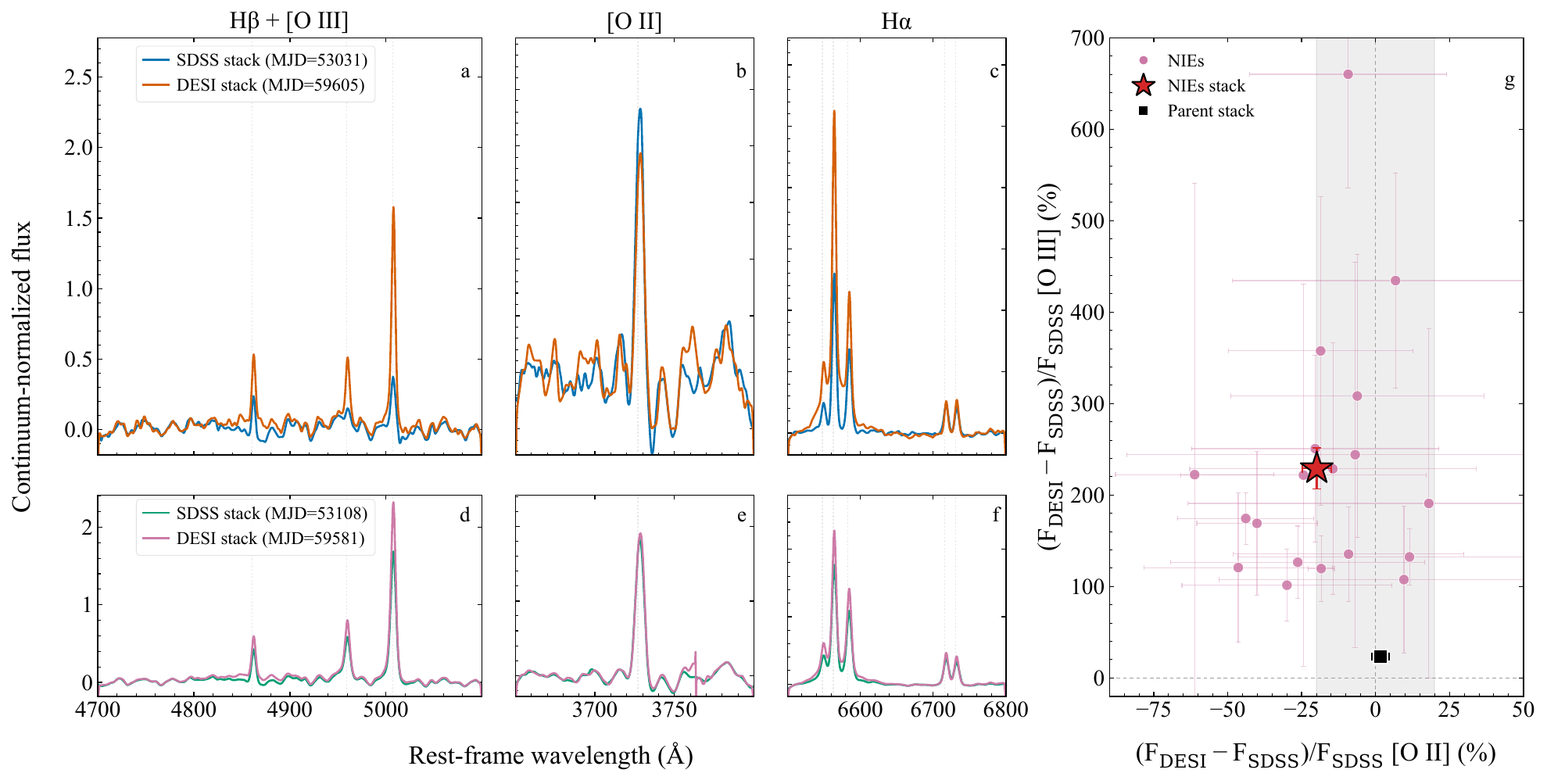}
\caption{
Selective [\ion{O}{3}] enhancement in the NIE hosts.
(a)--(c) Continuum-normalized stacked SDSS (blue) and DESI
(orange) spectra of the 19 NIE hosts around
H$\beta$+[\ion{O}{3}], [\ion{O}{2}], and
H$\alpha$+[\ion{N}{2}]+[\ion{S}{2}], respectively.
(d)--(f) Corresponding stacks for the 473 parent-sample nuclei,
constructed in the same spectral windows using the procedure described
in Appendix~\ref{subsec:systematic_tests}.
(g) Fractional [\ion{O}{2}] and [\ion{O}{3}] flux changes,
$(F_{\rm DESI}-F_{\rm SDSS})/F_{\rm SDSS}$, for the individual NIE
hosts; error bars represent $1\sigma$ uncertainties. The red star and
black square mark the NIE and parent stacks, respectively. The gray
band indicates
$|\Delta F_{\rm [O\,II]}/F_{\rm SDSS}|<20\%$.
The MJDs in the panel legends are the mean observation epochs of the
spectra contributing to each stack. The NIE hosts occupy the
high-[\ion{O}{3}]-response tail, whereas the parent stack shows only
weak average changes.
}
\label{fig:stacked_spectra}
\end{figure*}

% =====================================================================
% Table 2: Basic properties of the 19 NIE hosts
% =====================================================================

% \begin{rotatetable*}
\begin{deluxetable*}{lrrrrrrrrc}

\tabletypesize{\scriptsize}
\tablewidth{0pt}

\tablecaption{
Basic properties of the 19 NIE hosts.
\label{tab:nie_properties}
}

\tablehead{
\colhead{Name} &
\colhead{R.A.} &
\colhead{Decl.} &
\colhead{$z$} &
\colhead{SDSS MJD} &
\colhead{DESI MJD} &
\colhead{$\Delta t_{\rm rest}$} &
\colhead{$\Delta_{\rm [O\,III]}$} &
\colhead{$\log(M_{\rm BH}/M_{\odot})$} &
\colhead{{Catalog(s)}} \\
\colhead{} &
\colhead{(deg)} &
\colhead{(deg)} &
\colhead{} &
\colhead{} &
\colhead{} &
\colhead{(yr)} &
\colhead{(\%)} &
\colhead{} &
\colhead{}
}

\startdata
J0158
& 29.51980
& $-0.87270$
& 0.0807
& 52933
& 59569
& 16.8
& 120
& $6.07 \pm 0.18$
& {J21, Y25} \\
J0838
& 129.68470
& 35.16180
& 0.0645
& 52642
& 59601
& 17.9
& 358
& $6.89 \pm 0.15$
& {J21} \\
J1028
& 157.05640
& 2.78270
& 0.0713
& 52316
& 59623
& 18.7
& 174
& $7.75 \pm 0.10$
& {J21} \\
J1109
& 167.49310
& 37.13600
& 0.0260
& 53499
& 59666
& 16.5
& 435
& $8.06 \pm 0.06$
& {J21} \\
J1133
& 173.48304
& 67.01864
& 0.0397
& 51955
& 59710
& 20.4
& 132
& $8.07 \pm 0.07$
& {Y25} \\
J1149
& 177.34180
& 54.69760
& 0.0619
& 52706
& 59629
& 17.9
& 251
& $7.25 \pm 0.19$
& {J21} \\
J1300
& 195.10296
& $-2.48499$
& 0.1563
& 51694
& 59595
& 18.7
& 191
& $8.15 \pm 0.09$
& {Y25} \\
J1332
& 203.05260
& 20.61060
& 0.1125
& 54230
& 59716
& 13.5
& 108
& $7.45 \pm 0.15$
& {J21, N25} \\
J1342
& 205.68510
& 5.51560
& 0.0366
& 52373
& 59369
& 18.5
& 660
& $7.07 \pm 0.14$
& {J21} \\
J1438
& 219.55220
& $-1.51700$
& 0.1051
& 52409
& 59611
& 17.8
& 222
& $7.29 \pm 0.14$
& {J21} \\
J1448
& 222.12090
& 11.62560
& 0.0666
& 53531
& 59689
& 15.8
& 229
& $7.51 \pm 0.14$
& {J21} \\
J1513
& 228.44070
& 31.19030
& 0.0721
& 53149
& 59630
& 16.6
& 101
& $7.81 \pm 0.10$
& {J21, N25} \\
J1524
& 231.01080
& 5.91500
& 0.0973
& 54562
& 59369
& 12.0
& 135
& $6.92 \pm 0.21$
& {J21} \\
J1532
& 233.16508
& 44.53419
& 0.0375
& 52466
& 59385
& 18.3
& 169
& $5.78 \pm 0.14$
& {N25} \\
J1533
& 233.29180
& 27.48900
& 0.0721
& 54180
& 59684
& 14.1
& 127
& $8.04 \pm 0.08$
& {J21} \\
J1543
& 235.98998
& 10.10375
& 0.1754
& 54266
& 59683
& 12.6
& 222
& $7.80 \pm 0.13$
& {N25} \\
J1549
& 237.48000
& 33.46440
& 0.0857
& 53227
& 59716
& 16.4
& 121
& $6.72 \pm 0.22$
& {J21, N25} \\
J1620
& 245.14580
& 24.12400
& 0.0655
& 53226
& 59713
& 16.7
& 308
& $7.55 \pm 0.16$
& {J21, N25} \\
J2203
& 330.95520
& 11.40920
& 0.1862
& 52224
& 59527
& 16.9
& 244
& $7.99 \pm 0.13$
& {J21, Y25}
\enddata
\tablecomments{
R.A. and Decl. are given in decimal degrees.
The rest-frame spectroscopic baseline is calculated as
$\Delta t_{\rm rest}
=({\rm MJD}_{\rm DESI}-{\rm MJD}_{\rm SDSS})/
[365.25(1+z)]$.
The fractional [\ion{O}{3}] change is reported as
$\Delta_{\rm [O\,III]}
=(L_{\rm DESI}-L_{\rm SDSS})/L_{\rm SDSS}\times100\%$.
Black-hole masses are estimated from the SDSS stellar velocity
dispersion using the $M_{\rm BH}$--$\sigma_{\ast}$ relation of
\citet{Kormendy2013},
$\log(M_{\rm BH}/M_{\odot})
=8.49+4.38\log[\sigma_{\ast}/(200~{\rm km\,s^{-1}})]$.
The quoted black-hole mass uncertainties include the formal
uncertainty in $\sigma_{\ast}$ and the coefficient uncertainties of
the adopted relation.
{Catalogue memberships are denoted by J21
(\citealt{Jiang2021}), Y25 (\citealt{Yang2025}), and N25
(\citealt{Necker2025}); multiple labels indicate inclusion in more
than one input catalogue.}
}
\end{deluxetable*}

\section{Identification and Validation of Narrow-line Ionization Echoes}
\label{sec:nie_identification}

% We required both epochs to provide usable coverage of the principal
% diagnostic lines from
% [\ion{O}{2}]~$\lambda\lambda3726,3729$ to
% [\ion{S}{2}]~$\lambda\lambda6717,6731$. The rest-frame separation
% between the two spectroscopic epochs is defined as

% \begin{equation}
% \Delta t_{\rm rest}
% =
% \frac{{\rm MJD}_{\rm DESI}-{\rm MJD}_{\rm SDSS}}
% {365.25(1+z)},
% \label{eq:rest_baseline}
% \end{equation}

% yielding a characteristic baseline of approximately 16 yr for the
% parent sample. 

\subsection{Construction of the Two-epoch Parent Sample}
\label{subsec:parent_sample}

We constructed the two-epoch parent sample from three published
catalogues of galactic nuclei with evidence for recent nuclear
brightening: (i) MIR-variable galaxies identified among SDSS galaxies
\citep{Jiang2021}; (ii) turn-on changing-look AGN candidates
\citep{Yang2025}; and (iii) dust-echo-like infrared flares from the
Flaires catalogue \citep{Necker2025}. These catalogues represent
different observational manifestations of nuclear variability and
were used only to define the initial search list; they do not imply a
common physical triggering mechanism.
{Of the 19 NIEs identified below, 15 are included in
the \citet{Jiang2021} catalogue, with two additional sources each
from \citet{Yang2025} and \citet{Necker2025}. Two and four of the
Jiang et al.\ sources also appear in the Yang et al.\ and
Necker et al.\ catalogues, respectively; all catalogue memberships
are listed in Table~\ref{tab:nie_properties}.}

We combined the three catalogues, removed duplicate entries, and
required each source to have an earlier SDSS spectrum and a later DESI
spectrum. {No additional minimum SDSS--DESI time
separation was imposed.}
We imposed an upper redshift limit of $z<0.35$ and required
usable two-epoch coverage of the principal diagnostic lines from
[\ion{O}{2}]~$\lambda\lambda3726,3729$ through
[\ion{S}{2}]~$\lambda\lambda6717,6731$. These criteria yield a final
parent sample of 473 unique galactic nuclei.

No narrow-line flux, line ratio, or SDSS--DESI narrow-line change was
used in constructing the parent sample. The evidence used to select
the nuclear events is therefore independent of the spectroscopic
response investigated below. The parent sample serves both as the
search population for narrow-line ionization echoes and as an
empirical control for the ordinary two-epoch scatter introduced by
spectrophotometric calibration, fiber aperture, spectral resolution,
and intrinsic source variability.

Because the three input catalogues have different selection functions,
the resulting parent sample is neither volume limited nor a complete
census of nuclear transients. It is instead a variability-triggered
sample designed to test whether independently identified nuclear
brightenings are followed by a measurable and ionization-dependent
change in the optical narrow-line spectrum.

\subsection{Spectral Decomposition and Emission-line Measurements}
\label{subsec:spectral_decomposition}

We measured the emission-line fluxes using a uniform spectral-fitting
procedure applied independently to the SDSS and DESI spectra. The
continuum was modeled with the penalized pixel-fitting code
\texttt{pPXF}
\citep{Cappellari2004,Cappellari2017}, using host-galaxy stellar
population templates \citep{Bruzual2003} and, {when favored
by the Bayesian information criterion (BIC) relative to a stellar-only
continuum model}, a
featureless power-law component to represent the AGN continuum.
Emission lines were subsequently fitted in the
continuum-subtracted spectra.

We modeled the principal optical emission lines
[\ion{O}{2}]~$\lambda\lambda3726,3729$, H$\beta$,
[\ion{O}{3}]~$\lambda\lambda4959,5007$,
[\ion{O}{1}]~$\lambda6300$, H$\alpha$,
[\ion{N}{2}]~$\lambda\lambda6548,6584$, and
[\ion{S}{2}]~$\lambda\lambda6717,6731$ with Gaussian profiles.
The baseline model contained one narrow component for each line. The
[\ion{O}{3}]~$\lambda5007/\lambda4959$ and
[\ion{N}{2}]~$\lambda6584/\lambda6548$ flux ratios were fixed to 3.
The two components of the [\ion{O}{2}] doublet were assigned tied
kinematics, and their summed flux was used throughout the analysis.

The kinematics of narrow H$\beta$ and the core
[\ion{O}{3}] components were tied within the
H$\beta$+[\ion{O}{3}] fitting complex. Similarly, the kinematics of
[\ion{O}{1}], narrow H$\alpha$, [\ion{N}{2}], and
[\ion{S}{2}] were tied within the red-side
H$\alpha$+[\ion{N}{2}]+[\ion{S}{2}] complex. This approach reduces
degeneracies among neighboring lines while allowing the blue- and
red-side complexes to have independent narrow-line kinematics.

Additional spectral components were included only when favored by the
Bayesian information criterion. The candidate components included
optical Fe~\textsc{ii} emission, broad H$\alpha$ or H$\beta$, and a
second [\ion{O}{3}] component.
{The Fe~\textsc{ii} component was modeled with the
empirical I~Zw~1 template of \citet{Boroson1992}.}
When two [\ion{O}{3}] components were
selected, the reported [\ion{O}{3}] flux was the sum of the fitted
components. Among the 19 sources subsequently identified as NIE hosts,
six DESI spectra required weak broad Balmer components; these
components were modeled explicitly to prevent contamination of the
narrow-line measurements.

Throughout this work, [\ion{O}{2}] denotes the summed
$\lambda3726+\lambda3729$ flux,
[\ion{O}{3}] denotes $\lambda5007$,
[\ion{N}{2}] denotes $\lambda6584$, and
[\ion{S}{2}] denotes the summed
$\lambda6717+\lambda6731$ flux. Formal line-flux uncertainties
were propagated into the line-ratio and two-epoch luminosity-change
uncertainties. {An illustrative} spectral decomposition is shown in
Figure~\ref{fig:spectral_decomposition}, and the two-epoch narrow-line
luminosities of the NIE hosts are reported in
Table~\ref{tab:line_luminosities}.

We also measured the host-galaxy stellar velocity dispersion,
$\sigma_\ast$, from the \texttt{pPXF} continuum fit. Black-hole masses
were estimated using the local $M_{\rm BH}$--$\sigma_\ast$ relation
of \citet{Kormendy2013}. These masses are used for sample
characterization and for the model-dependent Eddington-ratio estimates
presented later; they do not enter the selection of the parent sample
or the operational definition of an NIE.

\subsection{Operational Definition of Narrow-line Ionization Echoes}
\label{subsec:nie_definition}

For each source, we quantified the two-epoch change in the fitted
narrow [\ion{O}{3}]~$\lambda5007$ flux as:

\[
\Delta_{\rm [O\,III]}
\equiv
\frac{
F_{\rm DESI}({\rm [O\,III]})
-
F_{\rm SDSS}({\rm [O\,III]})
}{
F_{\rm SDSS}({\rm [O\,III]})
}.
% \label{eq:oiii_fractional_change}
\]

We operationally define a NIE host as a
source satisfying:

\begin{equation}
\Delta_{\rm [O\,III]}>1,
\qquad\mathrm{or}\qquad
\frac{F_{\rm DESI}({\rm [O\,III]})}
     {F_{\rm SDSS}({\rm [O\,III]})}>2.
\label{eq:nie_selection}
\end{equation}

Thus, the narrow [\ion{O}{3}]~$\lambda5007$ flux must increase by more
than a factor of two between the SDSS and DESI epochs. This conservative
empirical threshold {was chosen to isolate} the extreme high-[\ion{O}{3}]-variability
tail of the parent population and lies well beyond its ordinary
{measurement and cross-survey} scatter, as demonstrated in
Appendix~\ref{subsec:systematic_tests}. The threshold is not intended to
represent a physically sharp boundary between responsive and
non-responsive narrow-line gas.

Applying Equation~(\ref{eq:nie_selection}) to the uniformly fitted
two-epoch spectra identifies 19 NIE hosts among the 473 parent-sample
nuclei, corresponding to $4.0\%$ of this variability-selected sample.
This fraction should not be interpreted as the incidence of NIEs among
all galaxies, because the parent sample was preselected for evidence of
recent nuclear variability and combines catalogues with different
selection functions. The 19 hosts span redshifts
$z=0.0260$--$0.1862$ and rest-frame SDSS--DESI baselines of
$12.0$--$20.4$ yr. 
Their basic properties are listed in
Table~\ref{tab:nie_properties}, and their two-epoch spectra  and light curves are shown in
{Figures~\ref{fig:j1342}, \ref{fig:individual_spectra}, and \ref{fig:individual_light_curves}}. {An illustrative} example, J134244.42+053056.1, is shown in Figure~\ref{fig:j1342}, where the
long-term MIR variability and the SDSS--DESI spectral evolution
illustrate the temporal context of a  NIE event.

The operational selection uses only the change in the narrow
[\ion{O}{3}]~$\lambda5007$ flux. The behavior of
[\ion{O}{2}], the Balmer lines, [\ion{N}{2}], and [\ion{S}{2}],
the evolution in diagnostic-line-ratio space, the MIR light-curve
properties, and the proposed physical nature of the nuclear event do
not enter the selection. These quantities therefore provide
independent tests of whether the selected sources exhibit a genuine
ionization-dependent response rather than a cross-survey flux offset or
a uniform rescaling of the narrow-line spectrum in Section \ref{sec:ionization_evolution}.

We use the term NIE phenomenologically. It denotes a large change in
the spatially unresolved narrow-line spectrum associated with a recent
nuclear ionizing episode, rather than a specific class of galaxy or
triggering mechanism. A NIE may follow renewed accretion in a
previously active nucleus, a tidal disruption event (TDE) or TDE-like flare, or a nuclear
transient whose physical origin remains uncertain. It is also distinct
from the spatially resolved, kiloparsec-scale ionization echoes used to
reconstruct nuclear activity over $10^{4}$--$10^{5}$ yr. Here,
``echo'' refers to a delayed narrow-line response detected over a
decade-scale two-epoch baseline and the available spectroscopy does not
directly resolve a source-specific reverberation lag.

\subsection{An Ionization-dependent Narrow-line Response}
\label{subsec:line_response}

By construction, every NIE host exhibits a greater than factor-of-two
increase in narrow [\ion{O}{3}]~$\lambda5007$ flux. The strength of
the stacked [\ion{O}{3}] enhancement therefore characterizes the
selected population but does not, by itself, provide an independent
validation of the NIE interpretation. The critical test is whether
the other emission lines, none of which entered the selection, show a
differential response consistent with a change in the ionizing
radiation field.

In Figure~\ref{fig:stacked_spectra}, we constructed equal-weight mean stacked spectra for the SDSS and DESI epochs in the three diagnostic windows described in
Appendix~\ref{subsec:systematic_tests}. For the 19 NIE hosts, the
fractional line-flux change measured from the locally
continuum-normalized stacks is

\[
\frac{\Delta F_{\rm line}}{F_{\rm SDSS,line}}
=
\frac{F_{\rm DESI,line}-F_{\rm SDSS,line}}
     {F_{\rm SDSS,line}}.
% \label{eq:stacked_line_change}
\]

{Here $\Delta F_{\rm line}$ is the difference
between the two stacked line measurements; $\Delta_{\rm [O\,III]}$
in Section~\ref{subsec:nie_definition} is already the fractional
change for an individual source.}

The strongest change occurs in
[\ion{O}{3}]~$\lambda5007$, which increases by
$229\pm22\%$. The narrow Balmer lines show more moderate increases of
$41\pm1\%$ in H$\beta$ and $59\pm1\%$ in H$\alpha$. By contrast, the
low-ionization forbidden lines remain substantially less variable:
the summed [\ion{O}{2}] doublet changes by $-20\pm5\%$,
[\ion{N}{2}]~$\lambda6584$ by $+12\pm1\%$, and the summed
[\ion{S}{2}] doublet by $-2\pm2\%$
(Figure~\ref{fig:stacked_spectra}). The enhancement is therefore
concentrated in [\ion{O}{3}], rather than being shared uniformly by
the optical narrow-line spectrum.

The same analysis applied to the complete 473-source parent sample
produces much weaker mean changes. The parent stack changes by
$+23\pm3\%$ in [\ion{O}{3}], $+2\pm3\%$ in [\ion{O}{2}],
$+4\pm0.4\%$ in narrow H$\beta$, $+6\pm0.8\%$ in narrow
H$\alpha$, $+5\pm1\%$ in [\ion{N}{2}], and $+2\pm1\%$ in
[\ion{S}{2}]. Thus, nuclear variability selection alone does not
produce the factor-of-several [\ion{O}{3}] enhancement or the
pronounced separation between [\ion{O}{3}] and the lower-ionization
forbidden lines observed in the NIE hosts.

The result is insensitive to the stacking estimator. Repeating the
analysis with inverse-variance weighting gives an
[\ion{O}{3}] increase of $254\pm32\%$ and an
[\ion{O}{2}] change of $-30\pm4\%$, together with substantially
smaller changes in the Balmer and low-ionization lines (Figure~\ref{fig:ext_weighted_stack}). The
source-by-source spectra likewise show that the differential response
is not generated by only one or two extreme objects
(Figure~\ref{fig:individual_spectra}).  The detail flux calibration and aperture effects are depicted in  Appendix~\ref{subsec:systematic_tests}.

The relative behavior of [\ion{O}{2}] and [\ion{O}{3}] is especially
informative because the two lines trace different oxygen ionization
states. The strong enhancement of O$^{2+}$ emission, accompanied by a
weak or negative change in O$^{+}$ emission, is inconsistent with a
single multiplicative rescaling of the spectrum. Instead, it indicates
a redistribution of the oxygen ionization balance toward the higher
ionization state. The moderate Balmer-line response and near constancy
of [\ion{N}{2}] and [\ion{S}{2}] further show that the observed
evolution is not simply a uniform increase in the total amount of
line-emitting gas.

These results establish the NIE hosts as a distinct spectroscopic
response population: their unresolved narrow-line gas evolves toward
a higher ionization state over the SDSS--DESI baseline. The
spectroscopy alone does not uniquely determine the physical trigger,
the source-specific response delay, or the distance of the variable
gas from the central black hole. The photoionization analysis in
Section~\ref{sec:ionization_evolution} tests the required change in the radiation
field, while the independent MIR record examined in
Section~\ref{sec:infrared} connects the narrow-line response to the
preceding nuclear event and provides the principal temporal and spatial
constraints.

\begin{figure*}[t]
    \centering
    \includegraphics[width=\textwidth]{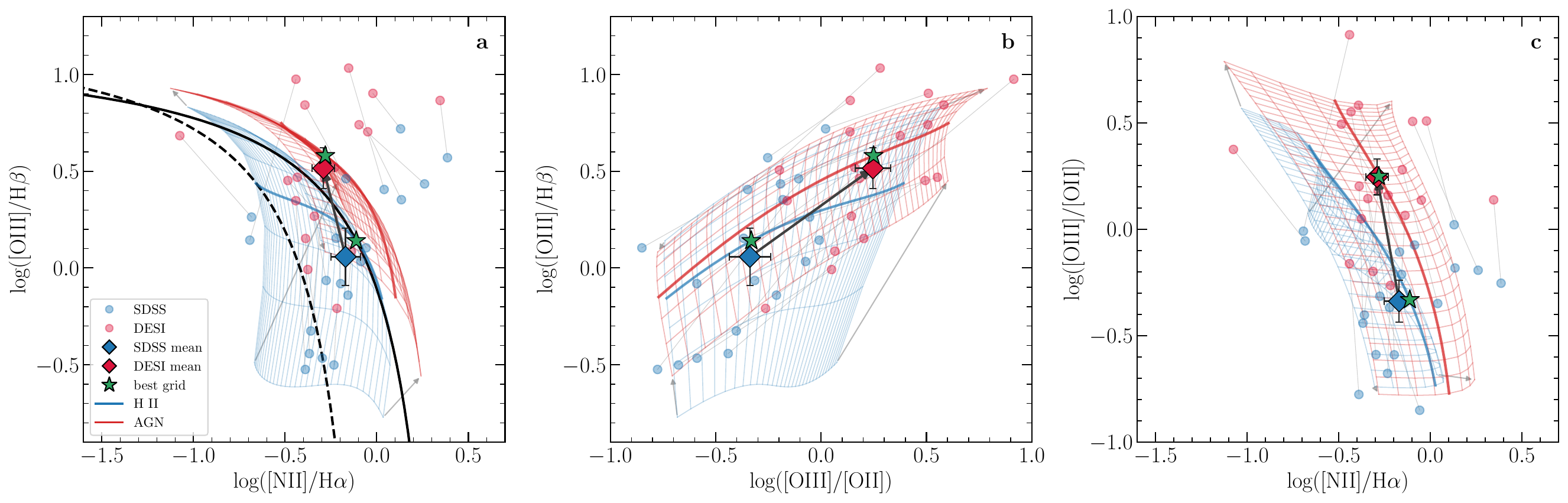}
    \caption{
    Coherent temporal evolution in narrow-line diagnostic diagrams.
    Blue and red circles represent measurements of individual NIE hosts
    at the SDSS and DESI epochs, respectively, with the two measurements
    for each source connected by a light-gray line.
    Large diamonds indicate the sample-averaged positions at the two
    epochs, and black arrows show the corresponding mean displacement
    from SDSS to DESI.
    The colored grids show representative photoionization models based
    on a stellar population spectral energy distribution (blue) and an
    AGN spectral energy distribution (red).
    Green stars mark the best-fitting grid points for the mean SDSS and
    DESI measurements.
    In panel (a), the black solid curve denotes the theoretical
    maximum-starburst boundary from \citet{Kewley2001}, while the black
    dashed curve denotes the empirical boundary from
    \citet{Kauffmann2003}.
    }
    \label{fig:oii_oiiii_evolution}
\end{figure*}

\begin{figure}[t]
    \centering
    \includegraphics[width=0.48\textwidth]{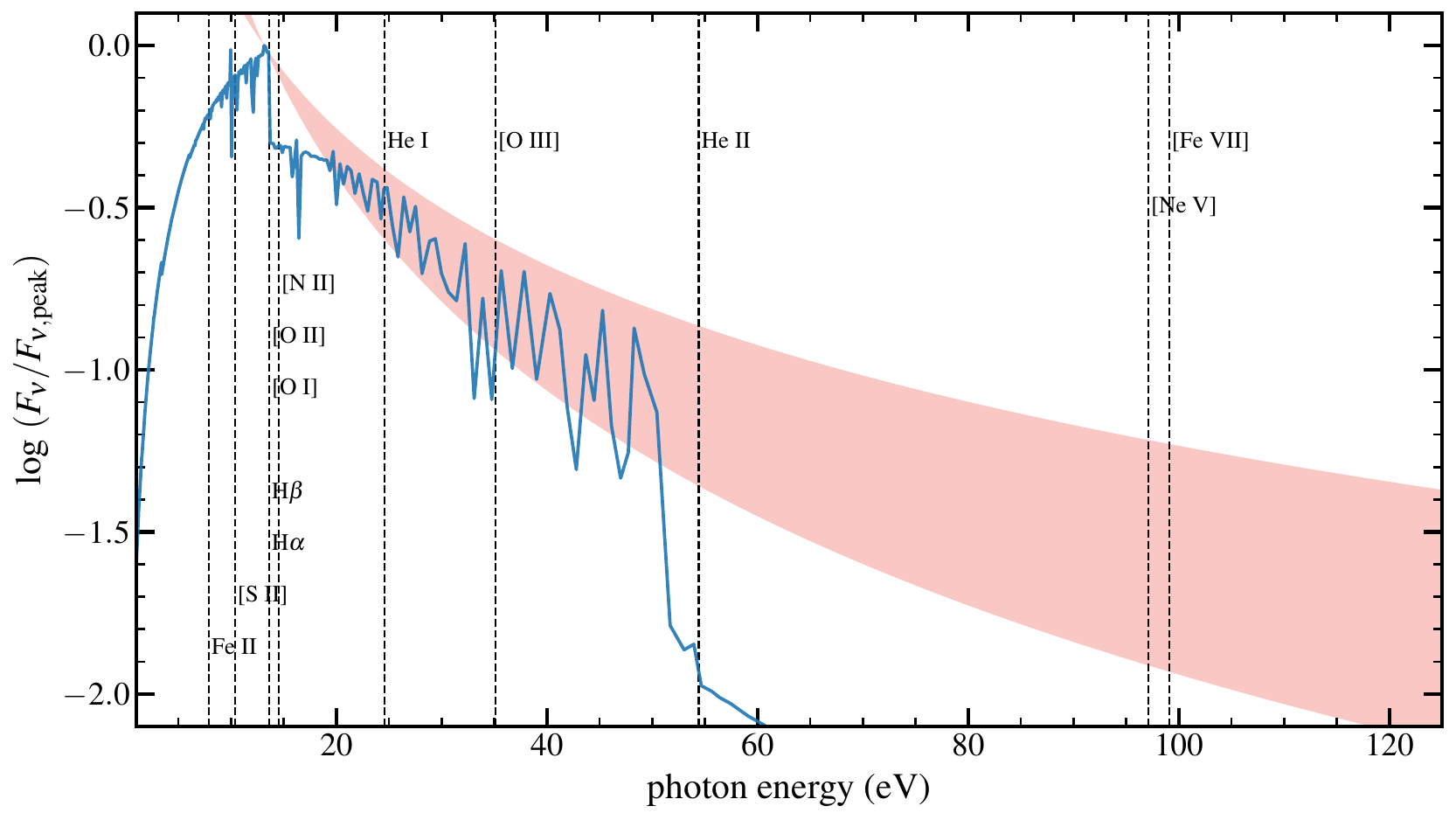}
    \caption{
    Spectral energy distributions adopted for the photoionization
    calculations.
    The incident spectra are shown as flux density per unit frequency
    and are normalized to the peak flux density of the stellar
    population SED.
    The red shaded region represents the range of AGN ionizing spectra
    with power-law indices from $\alpha=-2.2$ to $-1.4${,
    where $F_\nu\propto\nu^\alpha$ over
    $0.001\leq\lambda/\mu{\rm m}\leq0.25$}.
    The blue curve shows the stellar population SED {with
    stellar metallicity $Z_\star/Z_{\odot}=0.08$, selected from the
    \ion{H}{2}-region grids by the comparison described in
    Section~\ref{subsec:photoionization_constraints}}.
    Vertical dashed lines mark the ionization energies relevant to the
    ions producing the diagnostic emission lines.
    }
    \label{fig:SED}
\end{figure}

\section{Evolution of the Narrow-Line Ionization State}
\label{sec:ionization_evolution}

\subsection{Coherent Migration in Diagnostic Diagrams}
\label{subsec:diagnostic_evolution}

The selective enhancement of [O~{\sc iii}] is accompanied by coherent
evolution of the narrow-line ratios in multiple diagnostic diagrams.
To characterize this evolution, we define
\[
\begin{aligned}
{\rm O3} &\equiv
\log\left(
[{\rm O\,III}]\lambda5007/{\rm H}\beta
\right),\\
{\rm N2} &\equiv
\log\left(
[{\rm N\,II}]\lambda6584/{\rm H}\alpha
\right),\\
{\rm O32} &\equiv
\log\left(
[{\rm O\,III}]\lambda5007/
[{\rm O\,II}]\lambda\lambda3726,3729
\right).
\end{aligned}
\]
O3 traces the excitation of the higher-ionization narrow-line gas,
while O32 probes the relative populations of ${\rm O}^{+}$ and
${\rm O}^{2+}$ and is sensitive to both the ionization parameter and
the shape of the ionizing spectrum. N2 provides complementary
sensitivity to the gas-phase abundance, although it also depends on
the ionization state and the adopted abundance pattern. We therefore
interpret the three ratios jointly rather than treating any one of
them as a one-to-one diagnostic of a single physical parameter.

Figure~\ref{fig:oii_oiiii_evolution} shows the SDSS- and DESI-epoch
positions of the NIE hosts in the [N~{\sc ii}]-based BPT, O32--O3, and
O32--N2 planes. In the BPT plane, the sources migrate predominantly
toward higher O3 and toward the AGN locus. The mean N2 changes by only
\(-0.12\pm0.10\) dex, whereas the mean O3 increases by
\(0.46\pm0.18\) dex, with individual sources showing shifts approaching
1 dex. The nearly vertical displacement indicates that the transition
is driven primarily by an increase in excitation rather than by rapid
evolution along the abundance-sensitive direction.

The same behavior is seen more directly in the oxygen ionization
balance. The mean O32 increases by \(0.58\pm0.13\) dex between the
SDSS and DESI epochs. Because [O~{\sc ii}] and [O~{\sc iii}] trace
${\rm O}^{+}$ and ${\rm O}^{2+}$, respectively, the increase in O32
requires a larger incident flux of photons capable of ionizing
${\rm O}^{+}$ above its 35.12 eV ionization threshold. The coherent
increase in both O3 and O32, together with the comparatively weak
change in N2, therefore points to a change in the incident ionizing
radiation field rather than a uniform rescaling of all narrow
emission lines.

To quantify the physical origin of this displacement, we compare the
mean line-ratio positions at the two epochs with photoionization model
grids. The comparison tests whether the observed migration can be
reproduced by changes in gas-phase abundance, ionization parameter,
ionizing spectral shape, or a combination of these quantities.

\subsection{Photoionization Models and Parameter Constraints}
\label{subsec:photoionization_constraints}

To interpret the observed diagnostic-line migration, we constructed
H~{\sc ii}-region and AGN narrow-line-region photoionization grids
with MAPPINGS~V \citep{Sutherland2018}. The H~{\sc ii}-region grids
use Stromlo/FSPS models based on the Stromlo Stellar Tracks and FSPS
population-synthesis calculations
\citep{Grasha2021,Conroy2009,Conroy2010,Li2024}. The narrow-line-region
grids use broken-power-law AGN ionizing spectra with a variable
ultraviolet spectral index \(\alpha\){, defined by
\(F_\nu\propto\nu^\alpha\) over
\(0.001\leq\lambda/\mu{\rm m}\leq0.25\)}
\citep[see Figure~\ref{fig:SED}]{Groves2004a}. The
gas-phase abundance pattern follows the scaling of
\citet{Nicholls2017} and is parameterized by
\(12+\log({\rm O/H})\).
The full grid settings, including the ranges in ionization parameter
and gas abundance, geometry, pressure prescription, and dust
treatment, are summarized in Table~\ref{tab:mappings_grids}. For each
\({\rm SED}_{\rm HII}\) or \({\rm SED}_{\rm AGN}\), we computed the
corresponding \((U_{\rm H},Z_{\rm gas})\) grid for comparison with the
observed diagnostic ratios.

The incident ionizing spectral shape, ionization parameter
\(U_{\rm H}\), and gas-phase abundance \(Z_{\rm gas}\) are the
principal parameters controlling the strong-line ratios considered
here \citep{Groves2004a,Zhu2023}. We therefore performed
the diagnostic comparison in the parameter space
\(\Theta=(U_{\rm H},{\rm SED},Z_{\rm gas})\). We assumed
\(Z_{\rm gas,DESI}=Z_{\rm gas,SDSS}\), because bulk chemical
enrichment of the narrow-line-emitting gas is not expected over the
SDSS--DESI baseline. The comparison used the [N~{\sc ii}]-based BPT,
O32--O3, and O32--N2 planes. The search proceeded as follows.

\begin{deluxetable*}{lll}
\tablecaption{
Main parameter settings of the H~{\sc ii} region and narrow-line region
photoionization grids.
\label{tab:mappings_grids}
}

\tablehead{
\colhead{Parameter} &
\colhead{H~{\sc ii} region grids} &
\colhead{Narrow-line region grids}
}

\startdata
Comparison target &
SDSS epoch &
DESI epoch \\
Ionizing source &
Stromlo/FSPS stellar ionizing spectra &
AGN ionizing spectra \\
SED parameters &
$Z_\star/Z_\odot=0.02,0.08,0.20,0.40,0.70,1.00,1.30,1.70,2.00,2.50$
&
$F_\nu\propto\nu^\alpha$ at $0.001\leq\lambda/\mu{\rm m}\leq0.25$, \\
 & & $F_\nu\propto\nu^{-0.5}$ at $0.25<\lambda/\mu{\rm m}\leq10.0$, \\
 & & $F_\nu\propto\nu^2$ at $\lambda/\mu{\rm m}>10.0$, \\
 & & $-2.2\leq\alpha\leq-1.4$, step 0.1 \\
Ionization parameter &
$-3.5\leq\log U\leq-2.0$, step 0.05
&
$-3.5\leq\log U\leq-2.0$, step 0.05
\\
Gas abundance &
$12+\log({\rm O/H})=8.66-9.36$, step 0.1
&
$12+\log({\rm O/H})=8.66-9.36$, step 0.1
\\
Geometry &
Spherical &
Plane-parallel \\
Density structure &
Isobaric &
Isobaric \\
Initial pressure &
$\log(P/k)=5.5$
&
$\log(P/k)=7.0$
\\
Dust depletion &
Included &
Included \\
Dust destruction &
No &
No
\enddata
\tablecomments{
Dust depletion factors follow
\citet{Jenkins2009,Thomas2018}.
Iron depletion is fixed at 96.8\%,
corresponding to
$\log({\rm Fe}_{\rm free}/{\rm Fe}_{\rm total})=-1.5$.
}

\end{deluxetable*}

\begin{enumerate}

\item \textit{Selection of the initial \({\rm SED}_{\rm HII}\).}
We first selected the \({\rm SED}_{\rm HII}\) whose H~{\sc ii}-region
grid covered the SDSS-epoch line ratios in the [N~{\sc ii}]-based BPT
plane. The corresponding \((U_{\rm H},Z_{\rm gas})\) grid was adopted
as the initial reference grid for the SDSS epoch.

\item \textit{Determination of the candidate gas abundance.}
Within the selected \({\rm SED}_{\rm HII}\) grid, we compared the
SDSS-epoch mean position with the different gas-abundance sequences in
the BPT plane. The candidate \(Z_{\rm gas}\) was determined from the
abundance sequence with the minimum distance from the SDSS-epoch mean
position.

\item \textit{Determination of \({\rm SED}_{\rm AGN}\) at fixed
abundance.}
Keeping \(Z_{\rm gas}\) fixed, we compared the DESI-epoch mean position
with AGN narrow-line-region grids generated for different
\({\rm SED}_{\rm AGN}\). The adopted \({\rm SED}_{\rm AGN}\) was
selected by minimizing the distance between the DESI-epoch mean
position and the fixed-abundance sequence at the trial
\(Z_{\rm gas}\).

\item \textit{Determination of \(U_{\rm H}\) at the two epochs.}
The SDSS-epoch \(U_{\rm H}\) was determined by minimizing the distance
between the SDSS-epoch mean position and the grid points along the
selected abundance sequence in the \({\rm SED}_{\rm HII}\) grid. The
DESI-epoch \(U_{\rm H}\) was determined in the same manner from the
selected abundance sequence in the \({\rm SED}_{\rm AGN}\) grid.

\item \textit{Consistency check in the O32--N2 plane.}
For the resulting parameter combination
\((Z_{\rm gas},{\rm SED}_{\rm HII},{\rm SED}_{\rm AGN},
U_{\rm H,SDSS},U_{\rm H,DESI})\), we projected the corresponding
SDSS- and DESI-epoch grid points onto the O32--N2 plane. This plane
was used as an independent consistency check of the abundance- and
ionization-sensitive displacement, rather than to redefine
\(Z_{\rm gas}\), the SED, or \(U_{\rm H}\).

\item \textit{Ordered search over candidate SED combinations.}
If a parameter combination failed to reproduce the displacement
consistently across the three diagnostic planes, we first varied
\({\rm SED}_{\rm AGN}\) and repeated steps 3--5. If no candidate
\({\rm SED}_{\rm AGN}\) provided an acceptable solution, we varied the
initial \({\rm SED}_{\rm HII}\) and repeated the full procedure.

\end{enumerate}

This ordered search yielded a representative, self-consistent
parameter combination that reproduces the mean SDSS--DESI
displacement across the three diagnostic planes.

\subsection{Constraints on the Ionizing Spectrum and Ionization Parameter}
\label{subsec:physical_constraints}

We now use this representative solution to infer how the incident
ionizing radiation field changed between the SDSS and DESI epochs.

The representative solution has a matched gas-phase abundance of
\(12+\log({\rm O/H})=9.06\), corresponding to approximately
\(2\,Z_{\odot}\). The SDSS-epoch mean ratios are reproduced by an
H~{\sc ii}-region grid with a stellar ionizing spectrum characterized
by \(Z_{\star}/Z_{\odot}=0.08\) and
\(\log U_{\rm H}=-3.20\), whereas the DESI-epoch mean ratios are
reproduced by an AGN narrow-line-region grid with \(\alpha=-2.2\) and
\(\log U_{\rm H}=-2.65\)
(Figure~\ref{fig:oii_oiiii_evolution}).
The full range of DESI-epoch positions in the BPT plane requires AGN
continua extending from the representative value of \(\alpha=-2.2\)
to harder spectra with \(\alpha=-1.4\)
(see Appendix~\ref{app:agn_sed_range}). 

To quantify the change in spectral shape, we define the cumulative
ionizing-photon fraction
\[
\eta_E\equiv
\frac{Q(>E)}{Q(>13.6\,{\rm eV})},
\]
{Here \(Q(>E)\equiv\int_E^\infty
[L_{E'}/E']\,\mathrm{d}E'\) is the source photon emission rate above
\(E\) (\({\rm photons\,s^{-1}}\)); \(L_{E'}\) is the spectral luminosity
per unit energy.} We evaluate \(\eta_E\) at the
O\(^{+}\rightarrow\)O\(^{2+}\) threshold of \(35.12\,{\rm eV}\),
which sets the minimum photon energy required to produce
O\(^{2+}\), and at the
He\(^{+}\rightarrow\)He\(^{2+}\) threshold of \(54.42\,{\rm eV}\),
which serves as a reference for the harder extreme-ultraviolet tail.
For the adopted H~{\sc ii}-region spectrum,
\((\eta_{35},\eta_{54})=(12.0\%,0.87\%)\), whereas the representative
AGN continuum with \(\alpha=-2.2\) gives
\((\eta_{35},\eta_{54})=(12.4\%,4.7\%)\). The transition leaves
\(\eta_{35}\) nearly unchanged, but increases the relative
contribution of He\(^{+}\)-ionizing photons by a factor of
approximately 5.4. 

The representative solution also increases the ionization parameter
from \(\log U_{\rm H}=-3.20\) at the SDSS epoch to
\(\log U_{\rm H}=-2.65\) at the DESI epoch, corresponding to a factor
of approximately 3.5. For a given gas density of $n_\text{H}=10^3 \,\text{cm}^{-3}$, the incident photon
flux above energy \(E\) scales as
\[
\Phi_{>E}=n_{\rm H}c\,U_{\rm H}\eta_E.
\]
Combining the changes in \(U_{\rm H}\) and \(\eta_E\), the
representative model increases the photon flux above the
O\(^{+}\rightarrow\)O\(^{2+}\) threshold by a factor of approximately
3.7 and the flux above the
He\(^{+}\rightarrow\)He\(^{2+}\) threshold by a factor of
approximately 19.

The observed line-ratio migration therefore requires both a larger
incident ionizing flux and a reshaped accretion-powered continuum.
The enhanced flux above the O\(^{+}\) ionization edge drives the
observed shift from O\(^{+}\) toward O\(^{2+}\), while the much larger
increase at higher energies demonstrates that the extreme-ultraviolet
tail of the incident spectrum has strengthened substantially.

\begin{figure*}[t]
\centering
\includegraphics[width=0.8\linewidth]{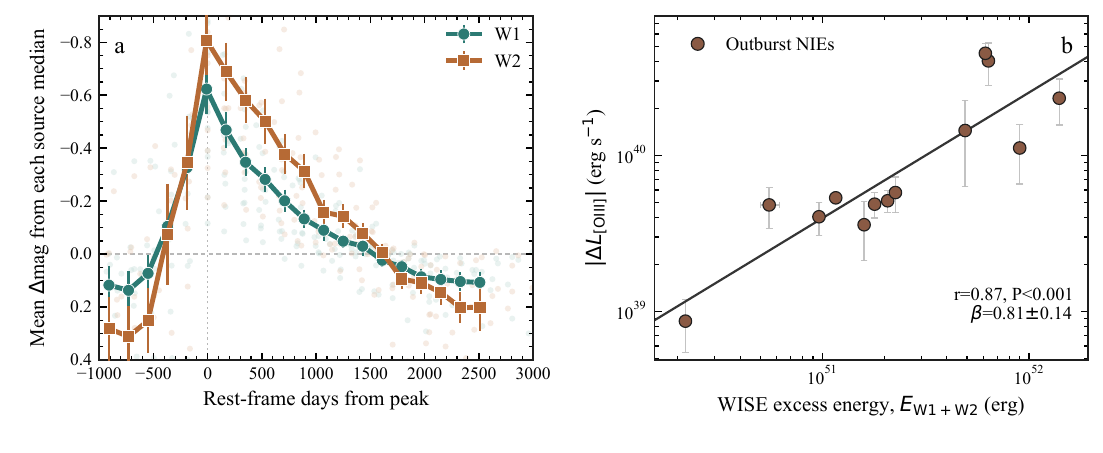}
\caption{
Mid-infrared variability and the narrow-line response of the outburst
NIEs.
(a) Rest-frame \textit{WISE} W1 (teal circles) and W2
(orange squares) light curves for the 13 NIE hosts with bounded
outburst windows. Both bands are aligned to the source-specific W1
peak and are shown as magnitude offsets from the median of each source.
Faint points show individual measurements, connected symbols show
source-equal-weighted averages in rest-frame time bins, and error bars
represent source-level uncertainties. The magnitude axis is inverted,
so upward excursions correspond to brightening.
(b) Integrated W1+W2 excess energy, $E_{\rm MIR}$, versus the absolute
change in [\ion{O}{3}] luminosity between the SDSS and DESI epochs for
the same 13 hosts; error bars represent $1\sigma$ uncertainties. The
black line shows the best-fitting log--log relation. The statistics
shown in the panel and the fitting procedure are described in
Section~\ref{subsec:mir_oiii_relation}. The six NIE hosts without
bounded outburst windows are included in
Figure~\ref{fig:mir_oiii_full19}.
}
\label{fig:mir_timescales}
\label{fig:mir_oiii_relation}
\label{fig:mir_oiii_outburst13}
\end{figure*}

\section{Mid-infrared Echoes and Their Connection to the Narrow-line Response}
\label{sec:infrared}

\subsection{Construction of the MIR Light Curves}
\label{subsec:wise_light_curves}

We constructed long-term MIR light curves using
single-exposure photometry from \textit{WISE} and the NEOWISE
reactivation mission. We used measurements in the W1 and W2 bands,
which trace nuclear radiation absorbed and reradiated by
circumnuclear dust. The MIR light curves therefore provide a record of
the nuclear brightening that is independent of the optical
narrow-line changes measured from the SDSS and DESI spectra.

For each source, the single-exposure measurements obtained during the
same observing period were grouped into a single photometric epoch.
The resulting epoch-binned W1 and W2 measurements were retained as a
function of the observer-frame Modified Julian Date (MJD). The SDSS
and DESI spectroscopic MJDs were recorded separately and marked on the
source-level light curves, allowing the two spectroscopic observations
to be placed within the longer-term infrared variability history.

For visualization and population-level comparisons, we removed the
source-to-source differences in absolute MIR magnitude by expressing
each light curve relative to its source-specific median. For each source
and each of the W1 and W2 bands, we define

\begin{equation}
\Delta m_b(t)
=
m_b(t)
-
\widetilde{m}_b,
\label{eq:wise_delta_mag}
\end{equation}

where \(b\) denotes the photometric band and
\(\widetilde{m}_b\) is the median epoch-binned magnitude of that source
in the corresponding band. With this definition,
\(\Delta m_b<0\) indicates MIR brightening. This normalization was
applied only when displaying and stacking the light curves. The
original epoch-binned photometry was retained for the flux-density
conversion, baseline subtraction, and excess-energy calculations.

The epoch-binned light curves were inspected to characterize the
long-term MIR behavior and to determine whether an outburst rise,
peak, and subsequent decline were sufficiently constrained. The
rest-frame peak alignment and population-level MIR timescales are
described in Section~\ref{subsec:mir_timescales}. The criteria used to
define the pre-outburst baseline and the bounded integration interval
are presented in Section~\ref{subsec:mir_energy}. Source-by-source W1
and W2 light curves, together with the available optical photometry and
the SDSS and DESI spectroscopic epochs, are shown in
{Figures~\ref{fig:j1342} and \ref{fig:individual_light_curves}}.

\subsection{Characteristic Mid-infrared Timescales}
\label{subsec:mir_timescales}

Among the 19 NIE hosts, 13 {satisfy the following
criteria in 100-day W2 bins before DESI: at least four bins, an observed peak
within the SDSS--DESI interval with $F_{\rm peak}/F_{\rm low}\geq1.5$,
and the first two bins satisfying
$F\leq F_{\rm low}+0.25(F_{\rm peak}-F_{\rm low})$.
Here $F_{\rm low}$ is the median of the faintest quartile of these
fluxes, using at least two bins.} We use these sources to
characterize the typical duration of the dust-reprocessed response.
The remaining six hosts {fail the requirement on the first two bins} and are therefore not included in the
fiducial timescale measurement. {A ``bounded'' outburst
here requires a measured low state preceding the rise, but not a full
return to baseline.}

For each of the 13 sources, we transformed the WISE observing times to
the source rest frame relative to the epoch of maximum W1 brightness: 
\begin{equation}
t_{\rm rest}
=
\frac{
{\rm MJD}-{\rm MJD}_{\rm W1,peak}
}{
1+z
}.
\label{eq:wise_rest_time}
\end{equation}

The W1 and W2 light curves were aligned using this source-specific
reference epoch. We retained measurements with
$t_{\rm rest}\geq-1000$ days in order to include the observed rise,
peak, and decline of the MIR outburst. As described in
Section~\ref{subsec:wise_light_curves}, each light curve was expressed
as a magnitude offset from its source-specific median from equation (\ref{eq:wise_delta_mag}), where negative values correspond to MIR brightening.

The aligned light curves were combined in uniform rest-frame time
bins. To prevent sources with more single-exposure measurements from
dominating the stack, all measurements from a given source within a
time bin were first averaged. The resulting source-level values were
then combined with equal weight. Uncertainties in the stacked light
curves were estimated from source-level resampling and the
source-to-source dispersion within each time bin. The resulting W1 and
W2 stacks are shown in Figure~\ref{fig:mir_timescales}.

We used the stacked W2 light curve to estimate two characteristic
timescales. The onset-to-half-decline timescale measures the interval
between the beginning of the MIR brightening and the stage at which the
outburst has declined to half of its peak excursion. The
onset-to-recovery timescale extends from the outburst onset to the
return of the stacked MIR signal to its baseline level. In the source
rest frame, these intervals are

\[
t_{\rm onset\mbox{-}half}
= 4.90~{\rm yr}, 
t_{\rm onset\mbox{-}rec}
= 6.55~{\rm yr}.
\]

For reference, the corresponding light-travel distances are
\[
c\,t_{\rm onset\mbox{-}half}
= 1.50~{\rm pc}, 
c\,t_{\rm onset\mbox{-}rec}
= 2.01~{\rm pc}.
\]

These values characterize the population-level duration of the
dust-reprocessed MIR signal. They should not be interpreted as direct
source-by-source reverberation lags or as unique measurements of the
dust radius, because the observed light curves also depend on the
duration of the illuminating event, the radial distribution of the
dust, and temporal smoothing across the emitting region. We instead
use them as empirical timescales for comparison with the
SDSS--DESI spectroscopic baselines in
Section~\ref{subsec:temporal_relation}.

\subsection{Temporal location constraint from the Mid-infrared Outbursts and Spectroscopic Baselines}
\label{subsec:temporal_relation}

\begin{figure*}[t]
\centering
\includegraphics[width=0.7\linewidth]
{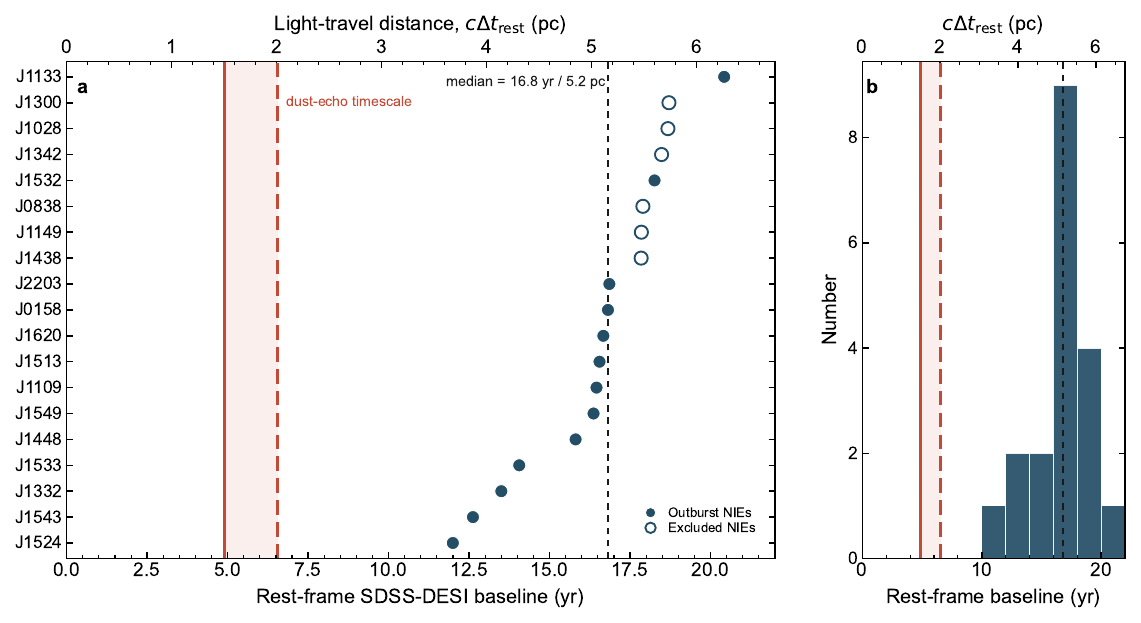}
\caption{
SDSS--DESI spectroscopic baselines compared with the characteristic
mid-infrared timescales.
(a) Rest-frame SDSS--DESI baselines for the 19 NIE hosts, ordered by
baseline. The upper axis gives the corresponding light-travel
distance, $c\Delta t_{\rm rest}$. Filled circles identify the 13 hosts
with bounded MIR outburst windows; open circles identify the six hosts
excluded from the regression in
Figure~\ref{fig:mir_oiii_relation}(b).
(b) Distribution of the same rest-frame baselines. The black dashed
line marks the sample median,
$\Delta t_{\rm rest}=16.8$ yr, corresponding to 5.2 pc in
light-travel distance. The shaded red interval shows the characteristic
MIR durations measured from the stacked W2 light curve. Its solid and
dashed edges mark the onset-to-half-decline interval of 4.90 yr and the
onset-to-recovery interval of 6.55 yr, corresponding to light-travel
distances of 1.50 and 2.01 pc, respectively. These intervals are
empirical duration constraints rather than direct reverberation-radius
measurements.
}
\label{fig:temporal_relation}
\label{fig:ext_restframe_baselines}
\end{figure*}

In Figure \ref{fig:ext_restframe_baselines}, the SDSS spectra provide the earlier narrow-line measurements and
predate the available WISE/NEOWISE monitoring, whereas the DESI
spectra provide the later narrow-line state after the observed MIR
activity had developed. For the 13 NIE hosts with bounded MIR
outbursts, the WISE light curves therefore place the nuclear
brightening within the interval bracketed by the two spectroscopic
epochs. This temporal ordering motivates the comparison between the
MIR energy released before the DESI observation and the corresponding
SDSS--DESI change in narrow-line emission.

For the 19 NIE hosts, the rest-frame SDSS--DESI baselines span

\[
12.0~{\rm yr}
\leq
\Delta t_{\rm spec,rest}
\leq
20.4~{\rm yr},
\]

with a median value of
\[
\widetilde{\Delta t}_{\rm spec,rest}
=
16.8~{\rm yr}.
% \label{eq:median_spectroscopic_baseline}
\]

For reference, the median baseline corresponds to a light-travel
distance of
\[
c\,\widetilde{\Delta t}_{\rm spec,rest}
=
5.2~{\rm pc}.
\]

Figure~\ref{fig:temporal_relation} compares the individual
spectroscopic baselines with the characteristic MIR timescales derived
in Section~\ref{subsec:mir_timescales}. The onset-to-half-decline and
onset-to-recovery times measured from the stacked W2 light curve are
$4.90$ and $6.55$ yr, corresponding to light-travel distances of
$1.50$ and $2.01$ pc, respectively. Every NIE host has an
SDSS--DESI baseline longer than these characteristic MIR intervals,
typically by a factor of a few. The earlier SDSS spectrum therefore
samples the nucleus before the later narrow-line state, while the
WISE/NEOWISE light curve records the intervening dust-reprocessed
activity.

The SDSS--DESI baseline should not be interpreted as a direct
measurement of the narrow-line reverberation lag. With only two
spectroscopic epochs, the time at which the narrow-line response began
is not resolved. The observed interval includes the unknown duration
of the ionizing episode, the light-travel delay to the emitting gas,
the gas response and recombination times, and any temporal smoothing
caused by an extended distribution of emitting material. Consequently, $c\,\Delta t_{\rm spec,rest}$ provides an observational upper bound on the characteristic
light-travel scale associated with the transition between the two
measured narrow-line states, rather than a direct radius measurement.

Likewise, the MIR onset-to-decline and onset-to-recovery intervals are
not direct dust reverberation lags. They reflect the convolution of the
intrinsic nuclear light curve with the spatial and thermal response of
the dust. We therefore use the MIR and spectroscopic intervals as
complementary empirical constraints: the MIR light curves identify and
characterize the intervening nuclear event, while the two
spectroscopic epochs bracket the associated change in the narrow-line
spectrum.  Their physical implications for the location of the
responsive gas are discussed in
Section~\ref{sec:discussion}.

\subsection{Integrated Mid-infrared Excess Energy}
\label{subsec:mir_energy}
To quantify the dust-reprocessed output associated with each nuclear
event, we measured the time-integrated MIR excess energy in the WISE
W1 and W2 bands. The calculation was performed using the original
epoch-binned photometry rather than the median-subtracted light curves
used for visualization and stacking.

The W1 and W2 Vega magnitudes were converted to flux densities using
the standard WISE zero-magnitude flux densities
\citep{Wright2010}. For band $b$, where
$b\in\{\mathrm{W1},\mathrm{W2}\}$, the flux density at time $t$ is:

\begin{equation}
F_{\nu,b}(t)
=
F_{\nu,0,b}\,10^{-0.4m_b(t)},
\label{eq:wise_flux_conversion}
\end{equation}

where $F_{\nu,0,b}$ is the zero-magnitude flux density of the
corresponding band.

For each source and band, we estimated a quiescent baseline flux
density, $F_{\nu,b}^{\rm base}$, {as the median of the
faintest quartile of the pre-outburst 100-day flux bins, using at least
two bins. When fewer than two pre-outburst bins were available, or no
outburst onset was identified, we applied the same estimator to all
bins preceding DESI.} The baseline-subtracted variable flux density
was then defined as:

\begin{equation}
\Delta F_{\nu,b}(t)
=
F_{\nu,b}(t)
-
F_{\nu,b}^{\rm base}.
\label{eq:wise_excess_flux}
\end{equation}

We converted the variable flux density into a baseline-subtracted band
luminosity using:

\begin{equation}
\Delta L_b(t)
\equiv
\Delta\!\left(\nu L_{\nu}\right)_b
=
4\pi D_{\rm L}^{2}\,
\nu_b\,
\Delta F_{\nu,b}(t),
\label{eq:wise_excess_luminosity}
\end{equation}

where $D_{\rm L}$ is the luminosity distance and $\nu_b$ is the
effective frequency of the WISE band. The combined W1+W2 variable
luminosity is:

\begin{equation}
\Delta L_{\rm W1+W2}(t)
=
\Delta L_{\rm W1}(t)
+
\Delta L_{\rm W2}(t).
\label{eq:wise_combined_luminosity}
\end{equation}

We define the integrated MIR excess energy as:

\begin{equation}
E_{\rm MIR}
=
\int_{t_{\rm onset}}^{t_{\rm DESI}}
\Delta L_{\rm W1+W2}(t)\,
{\rm d}t_{\rm rest},
\label{eq:mir_excess_energy}
\end{equation}

where $t_{\rm onset}$ is {the first W2 bin reaching
$F_{\rm low}+0.2(F_{\rm peak}-F_{\rm low})$, with $F_{\rm low}$ and
$F_{\rm peak}$ defined in Section~\ref{subsec:mir_timescales}},
$t_{\rm DESI}$ is the epoch of the DESI spectrum, and

\begin{equation}
{\rm d}t_{\rm rest}
=
\frac{{\rm d}t_{\rm obs}}{1+z}.
\label{eq:rest_frame_time_element}
\end{equation}

The upper integration limit is set by the DESI observation because
$E_{\rm MIR}$ is intended to quantify the dust-reprocessed radiation
released before the later narrow-line measurement. It therefore
provides the relevant preceding MIR record for comparison with the
SDSS--DESI change in [\ion{O}{3}].

The fiducial energy analysis was restricted to the 13 NIE hosts for
which the MIR rise and the excess-emission interval up to the DESI
epoch were sufficiently well constrained. For the remaining six
sources, the available WISE/NEOWISE light curves do not define a
bounded outburst interval, making the integrated excess energy
sensitive to the assumed onset or baseline. These sources are retained
in the spectroscopic and source-level light-curve analyses but are
excluded from regressions involving $E_{\rm MIR}$.

\begin{figure*}[t]
\centering
\includegraphics[width=0.8\linewidth]{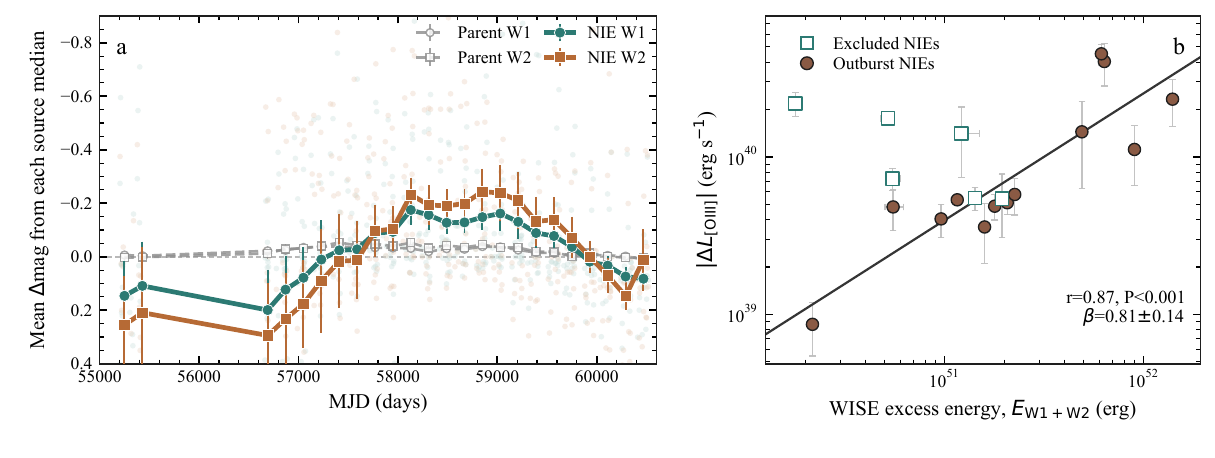}
\caption{
Mid-infrared variability and the
$E_{\rm MIR}$--[\ion{O}{3}] relation for the full NIE sample.
(a) Stacked \textit{WISE} W1 and W2 light curves as a function of MJD,
without peak alignment, for the 19 NIE hosts
(filled teal circles and orange squares) and the 473 parent-sample
nuclei (open gray symbols). Magnitudes are expressed as offsets from
the median of each source. Faint points show individual NIE
measurements, and error bars represent source-level uncertainties. The
magnitude axis is inverted, so upward excursions indicate brightening.
The NIE stacks show a pronounced MIR outburst, whereas the parent
stacks remain comparatively flat.
(b) Integrated W1+W2 excess energy, $E_{\rm MIR}$, versus the absolute
[\ion{O}{3}] luminosity change between the SDSS and DESI epochs;
error bars represent $1\sigma$ uncertainties. Filled circles identify
the 13 hosts with bounded outburst windows, and open squares identify
the six sources excluded from the regression. The black line and
reported statistics are derived from the 13 bounded-outburst hosts
only.
}
\label{fig:mir_oiii_full19}
\end{figure*}

\subsection{The Mid-infrared Energy--[O~{\sc iii}] Relation}
\label{subsec:mir_oiii_relation}

We next tested whether the strength of the delayed narrow-line response
is related to the dust-reprocessed output of the preceding nuclear
event. For each of the 13 NIE hosts with a bounded MIR outburst, we
compared the integrated W1+W2 excess energy,
$E_{\rm MIR}$, with the absolute change in narrow
[\ion{O}{3}]~$\lambda5007$ luminosity between the two spectroscopic
epochs,

\begin{equation}
\left|\Delta L_{\rm [O\,III]}\right|
\equiv
\left|
L_{\rm [O\,III],DESI}
-
L_{\rm [O\,III],SDSS}
\right|.
\label{eq:oiii_luminosity_change}
\end{equation}

The MIR energy was integrated only up to the DESI observation, as
defined in Equation~(\ref{eq:mir_excess_energy}). Thus,
$E_{\rm MIR}$ measures the dust-reprocessed radiation observed before
the later narrow-line luminosity measurement, preserving the temporal
ordering required for the comparison.

We fitted a linear relation in logarithmic space,

\begin{equation}
y = \alpha + \beta x,
\label{eq:mir_oiii_regression}
\end{equation}

where

\[
x =
\log_{10}
\left(
\frac{E_{\rm MIR}}{{\rm erg}}
\right), \\
y =
\log_{10}
\left(
\frac{
\left|\Delta L_{\rm [O\,III]}\right|
}{
{\rm erg\,s^{-1}}
}
\right).
\]

The two quantities are strongly correlated, with a correlation
coefficient of $r = 0.87$ and a probability of obtaining the observed correlation by chance of $P < 0.001$. The best-fitting logarithmic slope is $\beta = 0.81 \pm 0.14.$ The relation is shown in
Figure~\ref{fig:mir_oiii_relation}. The fitted slope is consistent
with an approximately proportional increase in the narrow-line
response with increasing MIR outburst energy: nuclei with more
energetic dust-reprocessed events generally exhibit larger changes in
[\ion{O}{3}] luminosity.

The six NIE hosts without bounded MIR outburst intervals are included
in the distribution of spectroscopic baselines but are distinguished
from the 13 fiducial outburst sources in
Figure~\ref{fig:mir_oiii_full19}. Their SDSS and DESI spectra remain
valid for the narrow-line analysis, but the incomplete MIR temporal
coverage prevents a reliable measurement of
$E_{\rm MIR}$ and therefore excludes them from the MIR-energy
regression. Their exclusion is therefore based on the temporal coverage of the MIR event rather than on their position relative to the fitted
relation.

The correlation provides an independent link between the MIR outburst
and the narrow-line evolution. The two quantities are derived from
different measurements and trace physically distinct components:
$E_{\rm MIR}$ records radiation absorbed and reradiated by
circumnuclear dust, whereas
$\left|\Delta L_{\rm [O\,III]}\right|$ measures the response of
ionized gas between the SDSS and DESI epochs. Their association
therefore supports the interpretation that both signals were driven by
the same enhanced nuclear radiation field.

We do not interpret Equation~(\ref{eq:mir_oiii_regression}) as a
calibrated conversion between MIR energy and [\ion{O}{3}] luminosity.
The measured $E_{\rm MIR}$ includes only the W1 and W2 excess and not
the full bolometric energy of the event. It also depends on the dust
covering factor, temperature distribution, viewing geometry, temporal
sampling, and the fraction of the nuclear radiation intercepted by the
dust. Similarly, the observed [\ion{O}{3}] change depends on the gas
density, covering factor, radial distribution, and response time.
The principal result is therefore the existence of a population-level
connection between two delayed records of the same nuclear episode,
rather than a universal MIR--[\ion{O}{3}] scaling law.

\begin{figure*}[t]
\centering
\includegraphics[width=\linewidth]{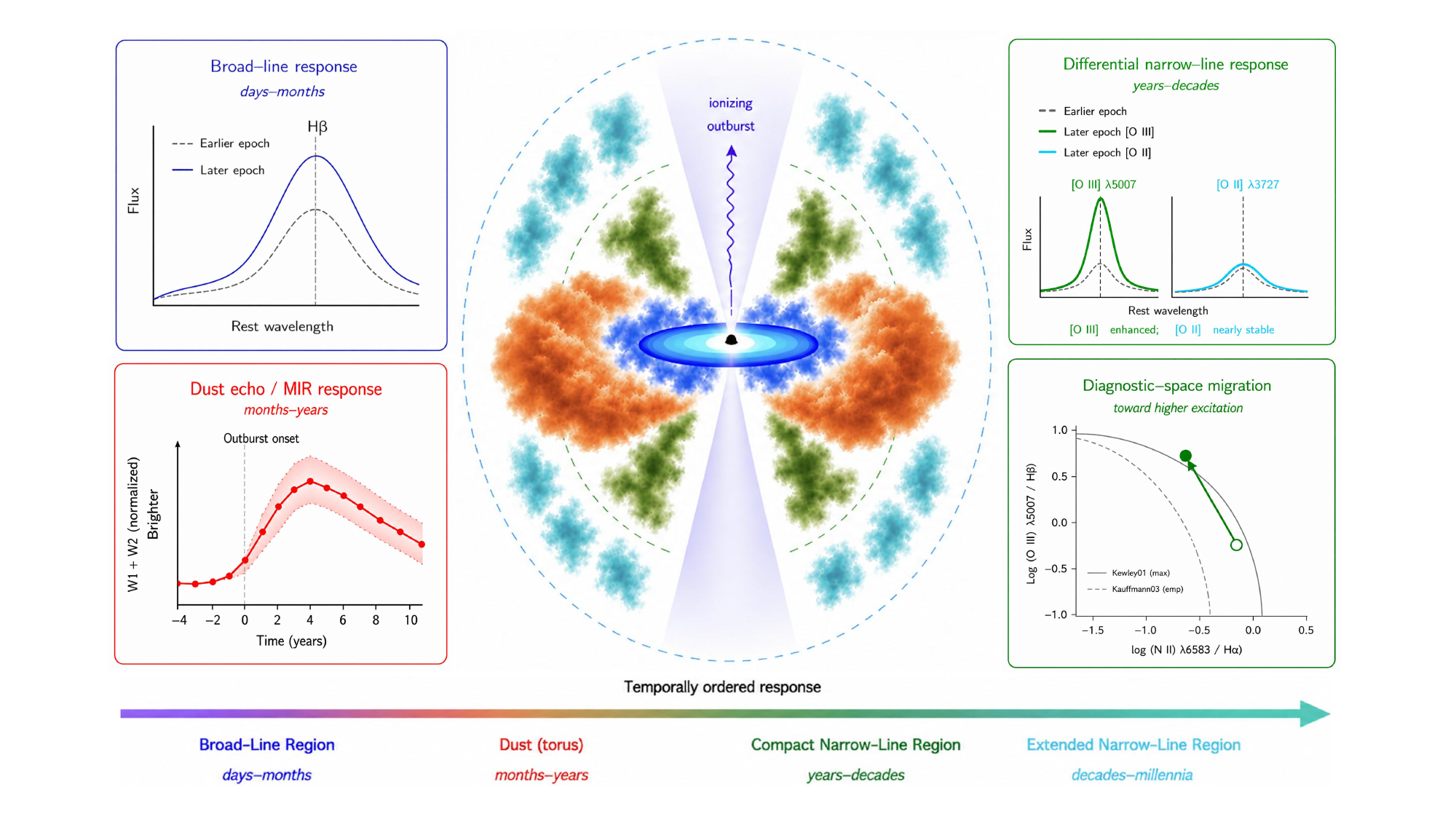}
\caption{
Ionization-memory hierarchy in galactic nuclei.
Schematic view of how a nuclear ionizing outburst is recorded by
different circumnuclear components. Broad-line emission traces the
immediate accretion change on days-to-months timescales, while dust
reprocessing records the same event as a mid-infrared echo on
months-to-years timescales. The NIEs reveal an intermediate,
years-to-decades layer in this hierarchy: compact, high-ionization
narrow-line gas on parsec scales can exhibit enhanced
[\ion{O}{3}] emission and coherent migration through optical
diagnostic diagrams. More extended or lower-ionization gas responds
over longer intervals and primarily preserves the longer-term nuclear
activity record. The schematic is not to scale.
}
\label{fig:ionization_memory}
\label{fig:schematic_hierarchy}
\end{figure*}

\section{Physical Interpretation and Implications}
\label{sec:discussion}

\subsection{A Compact Variable Narrow-line Component}
\label{subsec:compact_nlr}
The central implication of the NIEs is not that the entire classical
narrow-line region varies coherently on decade timescales. Rather, the
observations favor a compact and responsive narrow-line-emitting
component embedded within a more extended and comparatively stable
reservoir. In a persistently active nucleus, such gas would naturally
be described as the inner NLR. In a nucleus illuminated by a transient
event, however, it may instead represent pre-existing circumnuclear
clouds that become detectable only after the ionizing episode. The
NIE interpretation therefore does not require every host to have
contained a persistent AGN before the observed brightening.

Several aspects of the data support this picture. First, the later
DESI spectra show an absolute increase of more than a factor of two in
narrow [\ion{O}{3}]~$\lambda5007$, despite being obtained through a
smaller fiber than the earlier SDSS spectra. Second, the increase is
strongly dependent on ionization state: [\ion{O}{3}] changes much more
than [\ion{O}{2}], [\ion{N}{2}], and [\ion{S}{2}], while the narrow
Balmer lines show an intermediate response. Third, the sources evolve
coherently toward larger O3 and O32 in the diagnostic diagrams, whereas
the corresponding parent-sample controls show no comparable
population-wide migration. Finally, the [\ion{O}{3}] luminosity change
is related to the independently measured MIR excess energy. Together,
these observations favor an intrinsic change in the ionization state
of compact circumnuclear gas rather than a global rescaling of the
integrated narrow-line spectrum.

The SDSS--DESI baselines provide an upper timescale constraint on this
response. The 19 NIE hosts have rest-frame spectroscopic separations of
$12.0$--$20.4$ yr, with a median of $16.8$ yr, corresponding to
$5.2$ pc in light-travel distance. Because the nuclear event occurred
within the interval bracketed by the two spectra, the actual
light-travel delay between the ionizing source and the responsive gas
must generally be shorter than the full two-epoch separation. The
spectroscopic baseline therefore favors a characteristic scale of no
more than several parsecs, but it is not a direct measurement of the
gas radius.

The MIR light curves provide a complementary, but distinct, temporal
constraint. The stacked W2 light curve gives onset-to-half-decline and
onset-to-recovery intervals of $4.90$ and $6.55$ yr, respectively.
These timescales show that the dust-reprocessed signature of the
intervening nuclear event developed and declined well within the
SDSS--DESI interval. They should not, however, be interpreted as direct
dust reverberation lags or as lower limits on the
[\ion{O}{3}]-emitting radius. The observed MIR duration is a
convolution of the intrinsic event light curve with the spatial and
thermal response of the dust. The principal inference is therefore
temporal ordering: the MIR light curves identify an intervening
nuclear brightening, while the two spectroscopic epochs bracket the
associated narrow-line transition.

The response time of the gas also depends on density. Efficient
emission of the collisionally excited
[\ion{O}{3}]~$\lambda5007$ line requires an electron density below its
critical density,

\[
n_{\rm e}
\lesssim
n_{\rm crit}
\simeq
7\times10^{5}\ {\rm cm^{-3}},
\]

above which collisional de-excitation suppresses the line
\citep{Osterbrock2006}. At the same time, a response within the
observed interval favors gas dense enough that recombination and
ionization-equilibration times do not dominate over the light-travel
delay. The present spectra do not provide a unique density for the
variable component, and the observed transition is expected to
combine light-travel time, recombination time, the duration of the
ionizing episode, and geometric smoothing across the emitting
structure. The available data therefore constrain a region of
density--radius parameter space rather than providing a unique radius
or density measurement.

A two-component interpretation naturally explains the differing line
responses. The observed luminosity of a given narrow line can be
written schematically as
\[
L_{\rm line}
=
L_{\rm compact}
+
L_{\rm extended},
\]

where the compact component responds to the recent event and the
extended component averages the ionizing history over much longer
timescales. High-ionization [\ion{O}{3}] emission can receive a
substantial contribution from the responsive compact component,
whereas [\ion{O}{2}], [\ion{N}{2}], and [\ion{S}{2}] may remain
dominated by more extended gas. An increase in the ionizing flux can
also shift oxygen from O$^{+}$ to O$^{2+}$, enhancing
[\ion{O}{3}] without requiring a corresponding increase in
[\ion{O}{2}]. The observed decrease or weak evolution of
[\ion{O}{2}] is therefore qualitatively consistent with a change in
ionization balance, rather than an increase in the total mass of
line-emitting gas.

The fiber-aperture difference may affect the relative weights of these
components. Because the DESI fiber is smaller, it preferentially
samples compact nuclear emission while excluding a larger fraction of
the extended, slowly varying component. This could enhance the
contrast of an intrinsically variable compact component in the DESI
spectrum. It cannot, however, create the observed absolute increase in
[\ion{O}{3}] without genuine brightening of the nuclear line-emitting
gas, and the parent-sample tests already incorporate the ordinary
effects of the SDSS--DESI aperture difference. We therefore regard the
fiber change as a possible modifier of the measured variability
amplitude, but not as its primary origin.

The NIEs are consequently best interpreted as revealing a compact
high-ionization component within the unresolved narrow-line emission.
The data favor a characteristic location in the central few parsecs,
broadly comparable to the compact variable NLR inferred in NGC~5548
\citep{Peterson2013}, while allowing a more extended component to
remain effectively constant over the same interval. This layered
geometry explains how narrow lines can simultaneously act as
long-term records of nuclear activity and, under suitable conditions,
respond measurably to an individual recent ionizing episode.

\begin{figure}[t]
\centering
\includegraphics[width=\linewidth]
{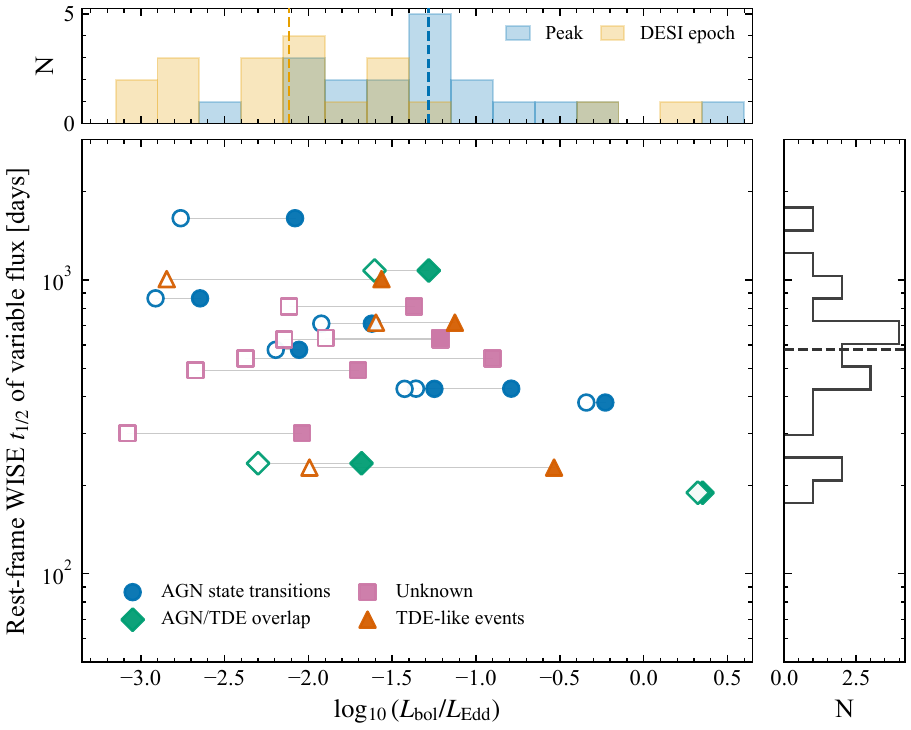}
\caption{
Eddington-ratio evolution, mid-infrared decline timescales, and
candidate triggers of the 19 NIE hosts.
Each source is shown at two epochs: filled symbols mark the
\textit{WISE} peak, open symbols mark the DESI epoch, and gray segments
connect the two. The Eddington ratio,
$\log_{10}(L_{\rm bol}/L_{\rm Edd})$, is estimated from the
baseline-subtracted variable MIR luminosity using the partial-covering
correction described in the text. Symbol shapes and colors identify
the candidate triggering classes inferred from published source-level
follow-up and the SDSS--DESI spectral properties.
The candidate AGN-state-transition systems are J0158, J1109, J1133,
J1300, J1332, J1513, J1532, J1533, J1543, and J2203
\citep{Frederick2019,Dai2020,Wang2022MIRONGII}.
The candidate TDE-like systems are J0158, J1342, J1513, J1549, J1620,
and J2203
\citep{Blanchard2017,Petrushevska2023,Jiang2025PS16dtm,
Sniegowska2026,Wang2012,Yang2013,Dou2016,He2021,Clark2024,
Onori2022AT2017gge,Zhou2026NeutrinoTDE}.
J0838, J1028, J1149, J1438, J1448, and J1524 are treated as ambiguous
nuclear transients. J0158, J1513, and J2203 occur in both the
AGN-transition and TDE-like groups and are plotted with the combined
AGN/TDE symbol. The vertical axis gives the measured rest-frame
half-decline time of the variable \textit{WISE} emission,
$t_{1/2}$. The upper and right panels show marginal distributions;
dashed lines mark the epoch-specific medians in the upper panel and the
sample median in the right panel.
}
\label{fig:wise_edd}
\end{figure}

\subsection{A Hierarchy of Ionization Memory}
\label{subsec:ionization_memory}

The NIEs add an intermediate layer to the sequence of delayed records
surrounding a variable galactic nucleus. A useful way to describe this
sequence is to treat each circumnuclear component as a response
function to the time-dependent ionizing luminosity. Schematically, the
luminosity of an emission component $j$ can be written as

\begin{equation}
L_j(t)
=
\int_{0}^{\infty}
\Psi_j(\tau)\,
Q_{\rm ion}(t-\tau)\,
{\rm d}\tau,
\label{eq:ionization_response_function}
\end{equation}

where $Q_{\rm ion}(t)$ represents the ionizing output of the nuclear
source and $\Psi_j(\tau)$ is the effective response function of the
gas or dust. The response function includes the spatial distribution,
covering factor, light-travel delay, density-dependent equilibration
time, and emissivity of the relevant component. Different regions
therefore preserve differently smoothed and delayed versions of the
same nuclear history.

When broad-line-emitting gas is present, the BLR provides the most
rapid response, typically on days-to-months timescales
\citep{Peterson2004,Du2018,Hu2021,Liyr2024}. Circumnuclear dust responds farther
outward, producing near- and mid-infrared echoes over months to years
\citep{Suganuma2006,Koshida2014,Lyu2019,Cackett2021}. The NIEs reveal
the next observable layer: compact narrow-line-emitting gas that can
retain a response over years to decades. At still larger radii, the
extended NLR and spatially resolved ionization echoes preserve nuclear
activity over much longer intervals, reaching
$10^{4}$--$10^{5}$ yr in kiloparsec-scale systems
\citep{Lintott2009,Keel2012}. The hierarchy is
illustrated schematically in Figure~\ref{fig:ionization_memory}.

The narrow-line memory is also ion dependent. The strong
[\ion{O}{3}] response, accompanied by much weaker changes in
[\ion{O}{2}], [\ion{N}{2}], and [\ion{S}{2}], indicates that the
different lines do not sample identical response functions. A compact
component exposed to the recent event can contribute substantially to
[\ion{O}{3}], while the lower-ionization lines remain dominated by
more extended gas that averages the radiation field over longer
timescales. Changes in the ionization parameter and spectral shape can
further redistribute oxygen from O$^{+}$ to O$^{2+}$, producing a
strong [\ion{O}{3}] response without a corresponding increase in
[\ion{O}{2}].

This picture should not be interpreted as a set of sharply separated,
concentric shells. Gas density, cloud covering factor, anisotropic
illumination, obscuration, and the intrinsic duration of the nuclear
event can produce overlapping response times at the same physical
radius. A high-density cloud at several parsecs may respond more
rapidly than lower-density gas closer to the nucleus, while an
extended distribution of clouds can broaden the observed response.
The hierarchy therefore describes the characteristic temporal
weighting of each observable rather than a unique radius assigned to
each emission line.

The MIR--[\ion{O}{3}] relation gives this hierarchy an empirical
connection across two components. The WISE excess records radiation
intercepted by dust, while the later [\ion{O}{3}] enhancement records
the response of ionized gas. Their correlation indicates that the two
memories scale with the strength of the same nuclear episode. In this
sense, NIEs bridge the gap between short-timescale reverberation and
the fossil record of the extended NLR, allowing ordinary narrow lines
to be used as time-domain observables rather than solely as static
classifiers.

\subsection{Possible Origins of the Ionizing Episodes}
\label{subsec:related_transients}
{Figure~\ref{fig:wise_edd} shows no clear separation among the proposed
changing-look, TDE-like, and ambiguous-transient groups in the
$\lambda_{\rm Edd}$--$t_{1/2}$ plane. Under the illustrative MIR-based
bolometric corrections adopted here, the median Eddington ratio
decreases from $\lambda_{\rm Edd}\sim0.04$ near the MIR peak to
$\sim0.007$ at the DESI epoch. This evolution crosses the
$\lambda_{\rm Edd}\sim0.01$ regime often associated with changing-look
or accretion-state transitions \citep{Noda2018,Guo2025CLAGN}.
However, these estimates depend on the dust covering factor and MIR
bolometric correction and cannot uniquely identify the trigger.}

{NIEs describe the delayed, ionization-dependent evolution of ordinary
optical narrow lines, rather than a specific triggering mechanism.
Changing-look AGNs are classified primarily through broad-line
appearance or disappearance, TDEs specify a physical fueling
mechanism, and ECLEs are defined by exceptionally strong coronal lines.
These categories are neither equivalent nor mutually exclusive, and
different triggers may produce similar narrow-line responses.}

{Both changes in a pre-existing accretion flow
\citep{Guo2025CLAGN,Guo2025b,Yang2025} and TDEs
\citep{Rees1988,Gezari2021,Blanchard2017} can supply the enhanced
ionizing radiation required for an NIE. The resulting dust and gas
echoes may outlast the directly observed continuum and broad-line
activity. Weak broad Balmer components are required in six DESI
spectra, but broad-line behavior was not used in the NIE selection.
Most hosts lack an unambiguous changing-look transition in the
available spectra. This does not exclude a major ionizing event,
because broad-line emission may be absent, obscured, diluted by the
host, or missed by the two-epoch sampling.}

{Variable ECLEs provide a close analogue through their evolution from
high-ionization coronal lines to lower-ionization features and, in
many cases, enhanced [\ion{O}{3}]
\citep{Yang2013,Clark2024,Short2023,Newsome2024,Hinkle2024,Kynoch2026}.
J1342 is a previously studied ECLE and long-duration TDE candidate
whose highest-ionization iron lines faded as lower-ionization features
emerged \citep{Wang2012,Yang2013,He2021,Clark2024}.
Its inclusion demonstrates that the NIE selection recovers an
established example of this sequence through the later
[\ion{O}{3}] response. Because NIEs do not require detectable coronal
emission, they may also select less extreme events, later evolutionary
stages, or gas with different density and ionization structure.
Persistent, AGN-like ECLEs further show that strong coronal emission
does not uniquely imply a TDE \citep{Clark2024,Kynoch2026}.}

{We therefore interpret NIEs as a common gas response to potentially
heterogeneous nuclear ionizing episodes. Likewise, movement toward
the AGN locus in diagnostic diagrams constrains the incident radiation
field but does not establish a pre-existing, long-lived AGN.
Identifying individual triggers requires early optical/UV and X-ray
observations, coronal and broad-line evolution, radio follow-up,
recurrence, and the persistence of the emission-line changes.}

\subsection{Incidence and Selection Effects}
\label{subsec:selection_effects}

Within the deliberately variability-selected parent sample, the
observed fraction of NIE hosts is $f_{\rm NIE|var}
=19/473.$ Thus, approximately $4\%$ of the nuclei selected for evidence of
recent brightening satisfy the adopted factor-of-two
[\ion{O}{3}] criterion. This fraction is conditional on the
time-domain selection of the parent population and should not be
interpreted as the fraction of all galaxies that host an NIE.

A separate estimate can be made using the MIR-variable galaxy
catalogue, from which 15 of the 19 NIE hosts were drawn. Of the 1026
MIR-variable sources in that catalogue, 359 have the DESI coverage
required for the present search, giving a DESI sampling fraction of $f_{\rm DESI}= 359/1026.$ Correcting the 15 detected NIEs for this sampling fraction and
normalizing by the $1{,}150{,}901$ galaxies searched in the original
MIR-variable parent population gives:

\begin{equation}
f_{\rm NIE,obs}
=
\frac{
N_{\rm NIE,MIR}
}{
f_{\rm DESI}\,N_{\rm gal}
}
\simeq
3.7\times10^{-5}.
\label{eq:nie_observed_fraction}
\end{equation}

Dividing by a characteristic rest-frame SDSS--DESI baseline of
approximately 17 yr gives the corresponding observed yield:

\begin{equation}
R_{\rm NIE,obs}
\simeq
\frac{f_{\rm NIE,obs}}{17~{\rm yr}}
\simeq
2.2\times10^{-6}\,
{\rm galaxy^{-1}\,yr^{-1}}.
\label{eq:nie_observed_rate}
\end{equation}

This calculation provides an order-of-magnitude occurrence rate for
large, detectable NIE transitions under the present selection. It is
not an intrinsic volumetric event rate. It assumes that an event
producing a detectable NIE occurs within the interval sampled by the
two spectra and that the resulting enhanced state is visible at the
DESI epoch. The calculation also does not separately account for the
duration of the ionizing event, the narrow-line response time, or the
visibility lifetime of the enhanced [\ion{O}{3}] emission.

Several selection effects influence both
Equations~\ref{eq:nie_observed_fraction} and
\ref{eq:nie_observed_rate}. First, the parent sample is enriched in
nuclei with detectable MIR variability, turn-on behavior, or
dust-echo-like flares. Events with weak dust emission, a small dust
covering factor, or MIR variability below the survey sensitivity may
therefore be missed. Conversely, the conditional $4\%$ fraction is
expected to be substantially larger than the incidence in an
unselected galaxy population.

Second, the NIE definition is intentionally conservative and
asymmetric. It selects only sources whose [\ion{O}{3}] flux increases
by more than a factor of two. Weaker responses, declining
[\ion{O}{3}] echoes, events dominated by other ionization states, and
sources whose response occurred entirely before or after the two
spectroscopic epochs are not included. The observed incidence
therefore characterizes only large positive transitions sampled at a
favorable stage of their evolution.

Third, inclusion requires usable SDSS and DESI spectra, appropriate
redshift coverage, and sufficiently reliable emission-line
measurements. DESI target selection, footprint, fiber assignment,
spectral signal-to-noise ratio, and the SDSS--DESI aperture difference
all affect the probability that an event enters the final sample.
Although the DESI sampling correction above accounts for one part of
this incompleteness, it does not constitute a complete selection
function.

Most of the current NIE hosts have estimated black-hole masses below
$10^{8}\,M_{\odot}$. Lower-mass systems may be favorable because a
given accretion episode can produce a larger fractional change in
luminosity and because the characteristic scale of the responsive gas
may be smaller. However, the present sample was not constructed to
measure a black-hole-mass-dependent NIE rate. The heterogeneous input
catalogues and small sample size prevent a robust demographic
inference. Establishing the intrinsic incidence as a function of
black-hole mass, host properties, event luminosity, and triggering
mechanism will require a larger sample with a forward-modeled
selection function.

\subsection{Future Observational Tests}
\label{subsec:future_tests}

{The priority is multi-epoch narrow-line spectroscopy of newly
identified nuclear outbursts, beginning before the expected response.
Repeated observations with the same instrument and aperture would
minimize cross-survey systematics and resolve whether [\ion{O}{3}]
strengthens, plateaus, or fades. Coordinated optical, ultraviolet,
X-ray, and MIR monitoring would constrain the incident continuum and
dust response, enabling direct measurements of narrow-line delays
and responsivities.}

{Integral-field spectroscopy could separate compact and extended
emission and test whether spatially resolved changes favor
photoionization over shocks or outflows. Density-sensitive doublets
and transauroral lines would constrain the responsive gas, while
He~\textsc{ii}, [\ion{Ne}{5}], and [\ion{Fe}{7}] would probe continuum
changes and connections with ECLEs. Velocity-resolved changes would
further constrain gas dynamics. Early optical/UV, X-ray, and radio
observations, together with broad- and coronal-line evolution, could
help distinguish accretion-flow changes from TDE-like flares.
NIE classification should remain independent of the underlying trigger.}

{Larger samples are needed to test the MIR--[\ion{O}{3}] relation
while accounting for its shared luminosity-distance dependence.
Partial-correlation tests, redshift-matched subsamples, flux- or
fluence-space comparisons, and hierarchical models incorporating
uncertainties and upper limits would address this issue. Repeated
spectra would also allow MIR energy to be compared with the evolving
line response.}

{Blind searches of repeat spectra, without prior MIR selection,
could uncover weaker or fading echoes and events in dust-poor
environments. Combining these searches with time-domain surveys
would permit detection-efficiency modeling and measurements of
NIE incidence across redshift, black-hole mass, host-galaxy
properties, event luminosity, and spectroscopic class.
Low-mass AGNs and intermediate-mass black-hole candidates are
promising targets \citep{Pucha2025}. Multi-epoch JWST spectroscopy
and infrared monitoring of little red dots could also test whether
analogous echoes occur at high redshift \citep{Greene2024,Lin2026}.}

\section{Conclusions}
\label{sec:conclusions}

We have presented a systematic search for decade-scale narrow-line
responses associated with recent nuclear variability by combining
WISE/NEOWISE mid-infrared light curves with repeat SDSS and DESI
spectroscopy. Our main conclusions are summarized as follows.

\begin{enumerate}

\item 
From a variability-selected parent sample of 473 galactic nuclei, we
identify 19 sources with more than a factor-of-two increase in narrow
[\ion{O}{3}]~$\lambda5007$ over rest-frame baselines of
$12.0$--$20.4$ yr. These NIEs show a
selective enhancement of [\ion{O}{3}], while lower-ionization forbidden
lines remain comparatively stable, indicating an ionization-dependent
response rather than a systematic flux offset.

\item 
The NIE hosts exhibit coherent evolution toward higher excitation in
optical diagnostic diagrams. The increase in
$\log([\mathrm{O\,III}]/\mathrm{H}\beta)$ and
$\log([\mathrm{O\,III}]/[\mathrm{O\,II}])$
indicates a substantial change in the ionization state of the
narrow-line gas. Photoionization modeling shows that this evolution is
consistent with an increase in the ionization parameter and a harder
ionizing spectral energy distribution, as expected when the gas is
illuminated by a recent nuclear outburst.

\item 
The independent MIR record links the narrow-line response to the same
nuclear event. For the 13 sources with well-constrained MIR outbursts,
the integrated W1+W2 excess energy correlates with the absolute
[\ion{O}{3}] luminosity change, suggesting that the dust echo and the
narrow-line response trace different stages of a common accretion
episode.

\item 
The NIEs reveal a compact, variable narrow-line component that provides
an intermediate ionization-memory layer between the rapid response of
the broad-line region, the infrared dust echo, and the long-lived
extended narrow-line region. The physical origin of the ionizing
episodes remains uncertain, with changing-look activity, TDE-like
events, and other nuclear transients all potentially contributing.
\end{enumerate}

The discovery of NIEs demonstrates that narrow emission lines are not
only static tracers of long-term nuclear activity, but can also preserve
the delayed imprint of individual accretion episodes on observable
timescales. By combining large-scale spectroscopic surveys with
time-domain infrared observations, it is now possible to reconstruct
the coupled evolution of the ionizing source, circumnuclear dust, and
ionized gas across previously inaccessible timescales. Future
wide-field repeat spectroscopy and multiwavelength monitoring will
determine whether NIEs represent a common phase of episodic
black-hole growth and establish how nuclear activity is transferred
from the accretion flow to the surrounding galactic environment.

\begin{acknowledgments}

{We thank the anonymous referee for helpful comments and suggestions that improved the clarity and presentation of this manuscript.}
We acknowledge support from National Key R\&D Program of China (2023YFA1607904 and 2021YFA1600404) and National Natural Science Foundation of China (NSFC) under grant Nos. 12503019, 12273041, and 12333003. LKM is grateful for support from a UKRI FLF [MR/Y020405/1]. We acknowledge the use of OpenAI's ChatGPT for language editing, manuscript-structure discussions and suggestions on the visual presentation of Figures; all scientific content, figures and conclusions remain the responsibility of the authors.

This material is based upon work supported by the U.S. Department of Energy (DOE), Office of Science, Office of High-Energy Physics, under Contract No. DE–AC02–05CH11231, and by the National Energy Research Scientific Computing Center, a DOE Office of Science User Facility under the same contract. Additional support for DESI was provided by the U.S. National Science Foundation (NSF), Division of Astronomical Sciences under Contract No. AST-0950945 to the NSF’s National Optical-Infrared Astronomy Research Laboratory; the Science and Technology Facilities Council of the United Kingdom; the Gordon and Betty Moore Foundation; the Heising-Simons Foundation; the French Alternative Energies and Atomic Energy Commission (CEA); the National Council of Humanities, Science and Technology of Mexico (CONAHCYT); the Ministry of Science, Innovation and Universities of Spain (MICIU/AEI/10.13039/501100011033), and by the DESI Member Institutions: \url{https://www.desi.lbl.gov/collaborating-institutions}.

Any opinions, findings, and conclusions or recommendations expressed in this material are those of the author(s) and do not necessarily reflect the views of the U. S. National Science Foundation, the U. S. Department of Energy, or any of the listed funding agencies.

The authors are honored to be permitted to conduct scientific research on Iolkam Du’ag (Kitt Peak), a mountain with particular significance to the Tohono O’odham Nation.

SDSS is managed by the Astrophysical Research Consortium for the Participating Institutions of the SDSS Collaboration including the Brazilian Participation Group, the Carnegie Institution for Science, Carnegie Mellon University, Center for Astrophysics | Harvard \& Smithsonian, the Chilean Participation Group, the French Participation Group, Instituto de Astrofísica de Canarias, The Johns Hopkins University, Kavli Institute for the Physics and Mathematics of the Universe(IPMU)/University of Tokyo, the Korean Participation Group, Lawrence Berkeley National Laboratory, Leibniz Institut fürAstrophysik Potsdam (AIP), Max-Planck-Institut für Astronomie (MPIA Heidelberg), Max-Planck-Institut für Astrophysik (MPA Garching), Max-Planck-Institut für ExtraterrestrischePhysik (MPE), National Astronomical Observatories of China, New Mexico State University, New York University, University of Notre Dame, Observatário Nacional/MCTI, The Ohio State University, Pennsylvania State University, Shanghai Astronomical Observatory, United Kingdom Participation Group, Universidad Nacional Autónoma de México, University of Arizona, University of Colorado Boulder, University of Oxford, University of Portsmouth, University of Utah, University of Virginia, University of Washington, University of Wisconsin, Vanderbilt University, and Yale University.

The PS1 has been made possible through contributions by the Institute for Astronomy, the University of Hawaii, the Pan-STARRS Project Office, the Max-Planck Society and its participating in- stitutes, the Max Planck Institute for Astronomy, Heidelberg and the Max Planck Institute for Extraterrestrial Physics, Garching, The Johns Hopkins University, Durham University, the University of Edinburgh, Queen’s University BELfast, the Harvard-Smithsonian Center for Astrophysics, the Las Cumbres Observatory Global Telescope Network Incorporated, the National Central University of Taiwan, the Space Telescope Science Institute, the National Aeronautics and Space Administration under Grant No. NNX08AR22G issued through the Planetary Science Division of the NASA Science Mission Directorate, the National Science Foundation under Grant No. AST-1238877, the University of Maryland, and Eotvos Lorand University (ELTE). ZTF is supported by the National Science Foundation under Grant No. AST-2034437 and a collaboration including Caltech, IPAC, the Weizmann Institute for Science, the Oskar Klein Center at Stockholm University, the University of Maryland, Deutsches Elektronen-Synchrotron and Humboldt University, the TANGO Consortium of Taiwan, the University of Wis- consin at Milwaukee, Trinity College Dublin, Lawrence Livermore National Laboratories, and IN2P3, France. Operations are conducted by COO, IPAC, and UW.

\end{acknowledgments}

\appendix

\section{Source-level Spectra, Light Curves, and Measurements}
\label{app:source_level}

This appendix presents the source-level material supporting the
population-level analysis in the main text. It includes {an
illustrative} spectral decomposition, the complete two-epoch spectra
and multiwavelength light curves of the 19 NIE hosts, and the
corresponding source properties and narrow-line luminosity
measurements.

Figure~\ref{fig:spectral_decomposition} shows {an illustrative}
spectral decomposition and illustrates how the narrow emission lines
were measured after modeling the host-galaxy continuum, nuclear
continuum, and broad Balmer components. The complete SDSS and DESI
spectra of the 19 NIE hosts are shown in
{Figures~\ref{fig:j1342} and \ref{fig:individual_spectra}}. These source-level spectra
demonstrate that the selective [\ion{O}{3}] enhancement is not driven
by only a few extreme objects.

{Figures~\ref{fig:j1342} and \ref{fig:individual_light_curves} present} the
\textit{WISE}/NEOWISE and available optical light curves, with the
SDSS and DESI spectroscopic epochs marked. Thirteen sources have
sufficiently bounded MIR outbursts for the timescale and excess-energy
analysis, while the remaining six retain valid spectroscopic NIE
classifications but are excluded from the $E_{\rm MIR}$ regression.

Table~\ref{tab:line_luminosities} provides
their fitted two-epoch narrow-line luminosities. These materials allow
the individual measurements underlying the stacked spectra,
diagnostic-line evolution, and MIR--[\ion{O}{3}] relation to be
examined directly.

\renewcommand{\thefigure}{A\arabic{figure}}
\setcounter{figure}{0}

\begin{figure*}[h]
\centering
\includegraphics[width=0.7\linewidth]{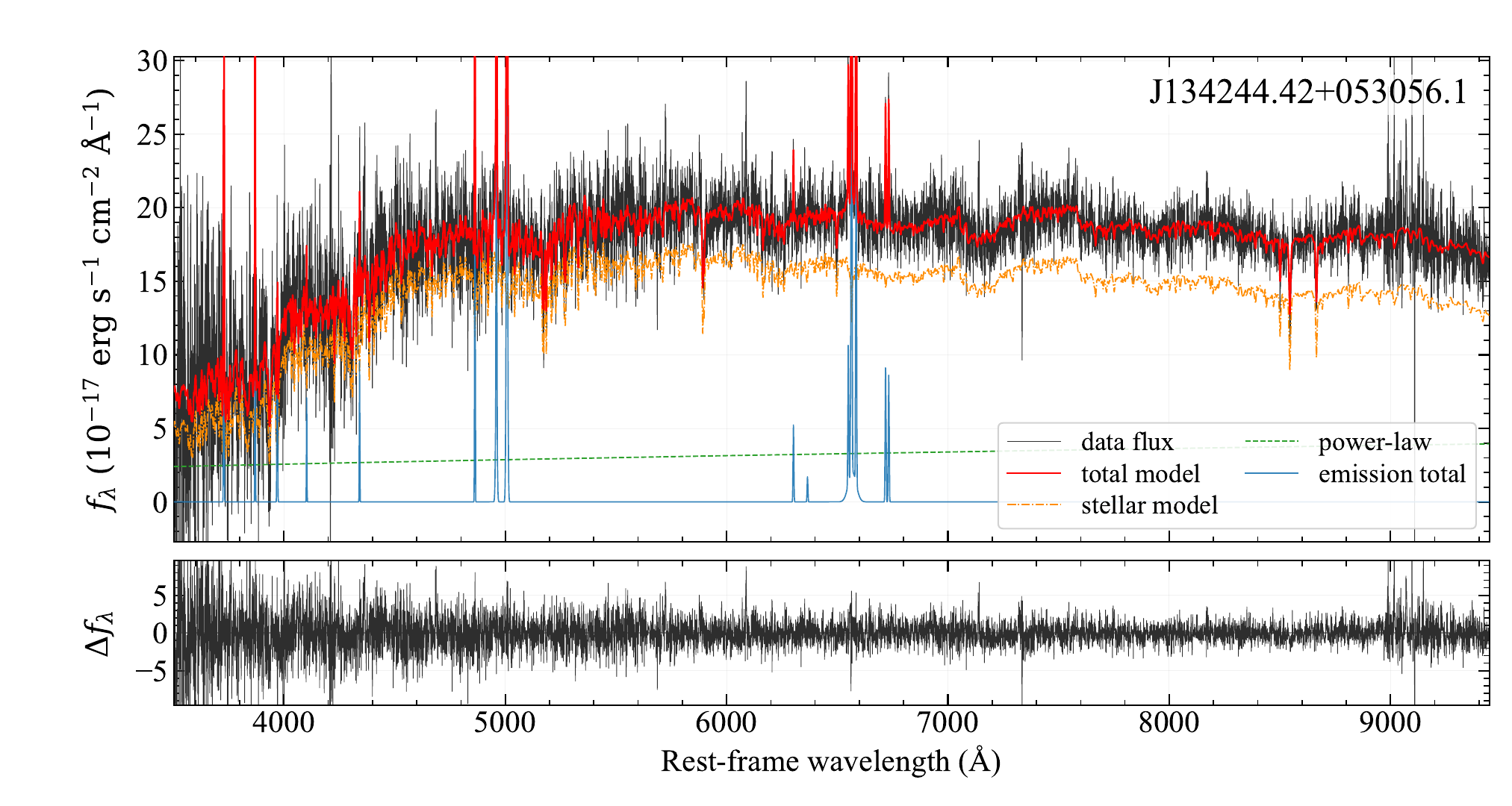}
\caption{
{Illustrative} spectral decomposition.
Example fit to the DESI spectrum of J134244.42+053056.1.
{The example uses the same source as Figure~\ref{fig:j1342}
to connect the decomposition with its spectral evolution. It is not
intended to represent the sample statistically.}
The {black} curve shows the observed spectrum and the
{red} curve shows the best-fitting total model. Colored curves show the principal fitted components,
including the host-galaxy stellar continuum, AGN power-law continuum,
emission-line model, and broad Balmer components where required. The
lower panel shows the residuals. This decomposition illustrates how
the narrow-line fluxes used in the two-epoch analysis were measured
after modeling the continuum and broad-line emission.
}
\label{fig:spectral_decomposition}
\label{fig:supp_specfit_example}
\end{figure*}

\begin{figure*}[p]
\centering
\includegraphics[width=0.93\linewidth]
{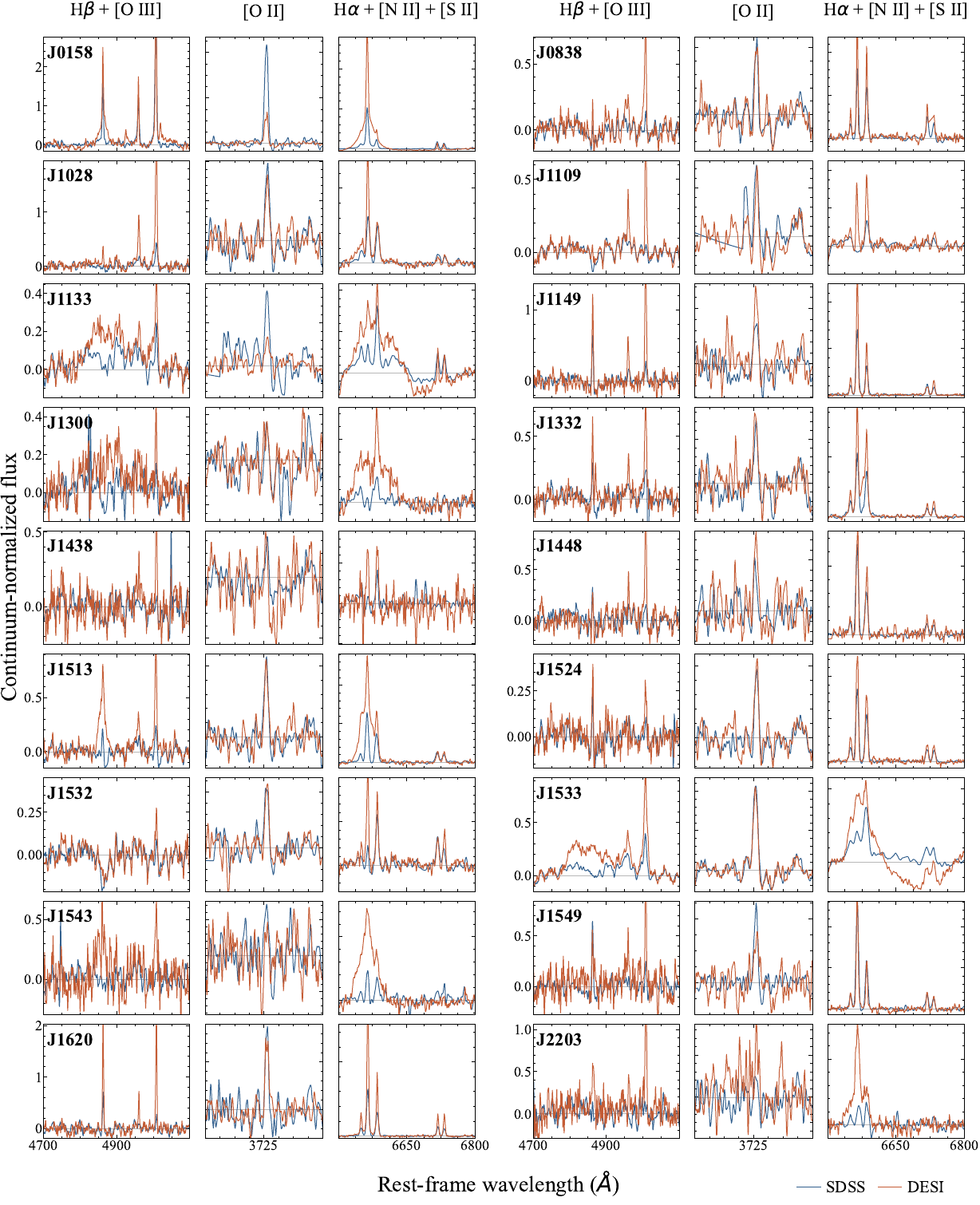}
\caption{
Full two-epoch spectra of the {remaining 18} NIE host galaxies.
{J1342 is shown in Figure~\ref{fig:j1342}.}
For each source, the three panels show the rest-frame
H$\beta$+[\ion{O}{3}], [\ion{O}{2}], and
H$\alpha$+[\ion{N}{2}]+[\ion{S}{2}] spectral windows.
Blue curves show the SDSS spectra and orange-red curves show the DESI
spectra. Across the sample, the H$\beta$+[\ion{O}{3}] region evolves
more strongly than the [\ion{O}{2}] window, while the
H$\alpha$+[\ion{N}{2}]+[\ion{S}{2}] region provides the corresponding
low-ionization comparison. The object-by-object spectra demonstrate
that the differential narrow-line evolution is not driven by only a
few extreme sources.
}
\label{fig:individual_spectra}
\label{fig:ext_fullspec}
\end{figure*}

\begin{figure*}[p]
\centering
\includegraphics[width=0.91\linewidth]{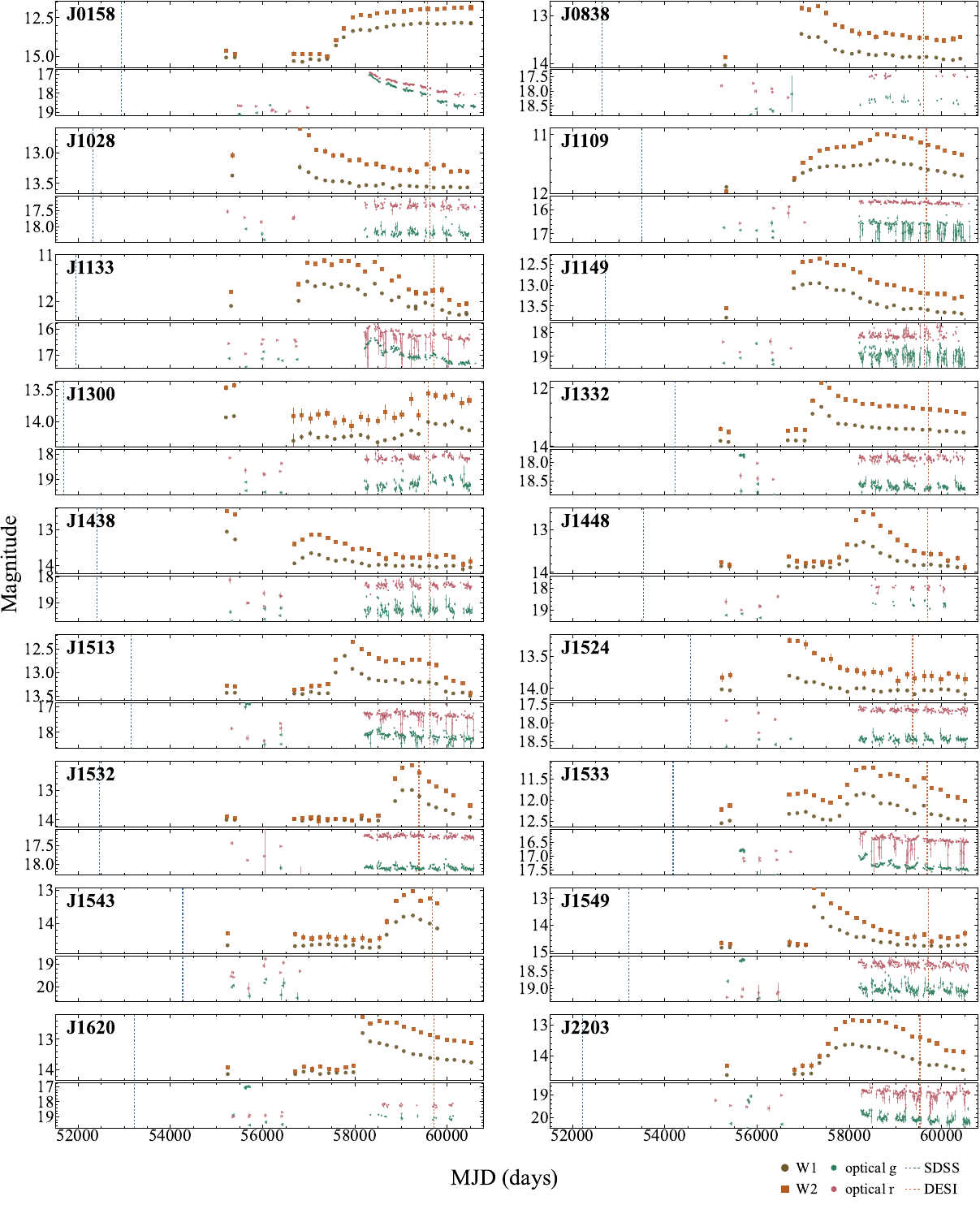}
\caption{
Source-by-source infrared and optical variability of the {remaining 18} NIE
hosts. {J1342 is shown in Figure~\ref{fig:j1342}.}
For each source, the upper panel shows the \textit{WISE} W1 and W2
light curves, and the lower panel shows the available optical light
curves. Vertical blue and orange-red dashed lines mark the SDSS and
DESI spectroscopic epochs, respectively. These light curves place the
two-epoch spectroscopy within the broader time-domain evolution of
each nucleus. {J0838, J1028, J1149, J1300, J1342, and
J1438 fail the initial-low-state criterion in
Section~\ref{subsec:mir_timescales} and are excluded from the fiducial
MIR timescale and excess-energy analyses.}
}
\label{fig:individual_light_curves}
\label{fig:supp_ir_optical}
\end{figure*}

\renewcommand{\thetable}{A\arabic{table}}
\setcounter{table}{0}

% =====================================================================
% Table 1: Two-epoch narrow-line luminosities
% =====================================================================

% \begin{longrotatetable}
\begin{deluxetable*}{llcccccc}

\tabletypesize{\scriptsize}
\tablewidth{0pt}

\tablecaption{
Two-epoch narrow-line luminosity measurements for the 19 NIE hosts.
\label{tab:line_luminosities}
}

\tablehead{
\colhead{Source} &
\colhead{Epoch} &
\colhead{[\ion{O}{2}]} &
\colhead{H$\beta$} &
\colhead{[\ion{O}{3}]} &
\colhead{H$\alpha$} &
\colhead{[\ion{N}{2}]} &
\colhead{[\ion{S}{2}]}
}

\startdata
J0158 & SDSS
& $40.63 \pm 0.02$
& $40.31 \pm 0.02$
& $40.58 \pm 0.07$
& $40.75 \pm 0.01$
& $40.07 \pm 0.02$
& $40.29 \pm 0.02$ \\
& DESI
& $40.54 \pm 0.02$
& $40.23 \pm 0.02$
& $40.92 \pm 0.02$
& $41.32 \pm 0.00$
& $40.25 \pm 0.04$
& $40.18 \pm 0.02$ \\[2pt]
J0838 & SDSS
& $39.90 \pm 0.02$
& $39.39 \pm 0.05$
& $39.31 \pm 0.15$
& $40.18 \pm 0.01$
& $39.99 \pm 0.01$
& $39.88 \pm 0.02$ \\
& DESI
& $39.81 \pm 0.16$
& $39.45 \pm 0.05$
& $39.97 \pm 0.04$
& $40.10 \pm 0.01$
& $39.87 \pm 0.01$
& $39.75 \pm 0.02$ \\[2pt]
J1028 & SDSS
& $40.11 \pm 0.05$
& $39.54 \pm 0.08$
& $40.00 \pm 0.04$
& $40.35 \pm 0.01$
& $40.18 \pm 0.02$
& $39.92 \pm 0.04$ \\
& DESI
& $39.86 \pm 0.17$
& $39.60 \pm 0.05$
& $40.44 \pm 0.01$
& $40.29 \pm 0.01$
& $39.90 \pm 0.02$
& $39.59 \pm 0.04$ \\[2pt]
J1109 & SDSS
& $39.28 \pm 0.22$
& $38.66 \pm 0.06$
& $39.09 \pm 0.09$
& $39.16 \pm 0.06$
& $39.42 \pm 0.05$
& $39.29 \pm 0.06$ \\
& DESI
& $39.31 \pm 0.06$
& $39.08 \pm 0.04$
& $39.82 \pm 0.02$
& $39.60 \pm 0.02$
& $39.50 \pm 0.03$
& $39.17 \pm 0.04$ \\[2pt]
J1133 & SDSS
& $39.75 \pm 0.11$
& $39.21 \pm 0.06$
& $39.57 \pm 0.04$
& $39.78 \pm 0.02$
& $39.92 \pm 0.02$
& $39.85 \pm 0.02$ \\
& DESI
& $39.79 \pm 0.20$
& $39.23 \pm 0.17$
& $39.93 \pm 0.04$
& $40.02 \pm 0.11$
& $39.97 \pm 0.12$
& $39.85 \pm 0.03$ \\[2pt]
J1149 & SDSS
& $39.78 \pm 0.13$
& $39.78 \pm 0.02$
& $39.34 \pm 0.12$
& $40.48 \pm 0.01$
& $40.12 \pm 0.01$
& $39.93 \pm 0.02$ \\
& DESI
& $39.68 \pm 0.19$
& $39.73 \pm 0.03$
& $39.89 \pm 0.04$
& $40.39 \pm 0.01$
& $40.00 \pm 0.01$
& $39.85 \pm 0.04$ \\[2pt]
J1300 & SDSS
& $40.12 \pm 0.27$
& $39.30 \pm 0.39$
& $39.87 \pm 0.27$
& $39.84 \pm 0.14$
& $40.23 \pm 0.07$
& $39.96 \pm 0.15$ \\
& DESI
& $40.19 \pm 0.12$
& $39.47 \pm 0.28$
& $40.33 \pm 0.10$
& $40.18 \pm 0.10$
& $40.53 \pm 0.07$
& $40.09 \pm 0.08$ \\[2pt]
J1332 & SDSS
& $40.23 \pm 0.07$
& $40.16 \pm 0.03$
& $40.02 \pm 0.16$
& $40.97 \pm 0.01$
& $40.81 \pm 0.01$
& $40.44 \pm 0.02$ \\
& DESI
& $40.27 \pm 0.24$
& $40.25 \pm 0.02$
& $40.33 \pm 0.05$
& $40.97 \pm 0.01$
& $40.84 \pm 0.01$
& $40.47 \pm 0.02$ \\[2pt]
J1342 & SDSS
& $39.53 \pm 0.13$
& $39.38 \pm 0.02$
& $39.52 \pm 0.03$
& $40.08 \pm 0.01$
& $39.38 \pm 0.02$
& $39.37 \pm 0.03$ \\
& DESI
& $39.49 \pm 0.10$
& $39.42 \pm 0.04$
& $40.40 \pm 0.06$
& $40.06 \pm 0.01$
& $39.62 \pm 0.02$
& $39.44 \pm 0.02$ \\[2pt]
J1438 & SDSS
& $39.74 \pm 0.13$
& $38.98 \pm 0.37$
& $39.39 \pm 0.26$
& $39.58 \pm 0.08$
& $39.62 \pm 0.11$
& $39.14 \pm 0.25$ \\
& DESI
& $39.62 \pm 0.20$
& $38.86 \pm 0.27$
& $39.90 \pm 0.10$
& $39.91 \pm 0.05$
& $39.76 \pm 0.09$
& $39.38 \pm 0.17$ \\[2pt]
J1448 & SDSS
& $39.65 \pm 0.08$
& $39.57 \pm 0.05$
& $39.25 \pm 0.18$
& $40.27 \pm 0.01$
& $39.91 \pm 0.02$
& $39.65 \pm 0.04$ \\
& DESI
& $39.58 \pm 0.23$
& $39.30 \pm 0.05$
& $39.76 \pm 0.05$
& $40.01 \pm 0.01$
& $39.73 \pm 0.02$
& $39.45 \pm 0.04$ \\[2pt]
J1513 & SDSS
& $40.07 \pm 0.04$
& $39.82 \pm 0.04$
& $39.76 \pm 0.07$
& $40.59 \pm 0.01$
& $40.32 \pm 0.01$
& $40.11 \pm 0.02$ \\
& DESI
& $39.92 \pm 0.22$
& $39.79 \pm 0.24$
& $40.06 \pm 0.04$
& $40.48 \pm 0.01$
& $40.14 \pm 0.02$
& $39.96 \pm 0.02$ \\[2pt]
J1524 & SDSS
& $40.23 \pm 0.17$
& $40.05 \pm 0.03$
& $39.55 \pm 0.07$
& $40.89 \pm 0.01$
& $40.66 \pm 0.01$
& $40.39 \pm 0.02$ \\
& DESI
& $40.19 \pm 0.07$
& $40.13 \pm 0.04$
& $39.92 \pm 0.07$
& $40.87 \pm 0.01$
& $40.65 \pm 0.01$
& $40.37 \pm 0.02$ \\[2pt]
J1532 & SDSS
& $39.56 \pm 0.09$
& $38.60 \pm 0.51$
& $38.71 \pm 0.08$
& $39.30 \pm 0.02$
& $39.24 \pm 0.03$
& $39.34 \pm 0.03$ \\
& DESI
& $39.34 \pm 0.12$
& $38.63 \pm 0.10$
& $39.14 \pm 0.10$
& $39.50 \pm 0.01$
& $39.19 \pm 0.02$
& $39.24 \pm 0.03$ \\[2pt]
J1533 & SDSS
& $40.48 \pm 0.25$
& $39.78 \pm 0.08$
& $40.50 \pm 0.02$
& $40.45 \pm 0.03$
& $40.58 \pm 0.02$
& $40.38 \pm 0.03$ \\
& DESI
& $40.35 \pm 0.04$
& $39.96 \pm 0.06$
& $40.86 \pm 0.07$
& $40.63 \pm 0.05$
& $40.61 \pm 0.07$
& $40.43 \pm 0.02$ \\[2pt]
J1543 & SDSS
& $40.18 \pm 0.12$
& $39.66 \pm 0.33$
& $39.81 \pm 0.42$
& $40.48 \pm 0.05$
& $40.25 \pm 0.05$
& $40.27 \pm 0.09$ \\
& DESI
& $39.77 \pm 0.28$
& $39.85 \pm 0.07$
& $40.32 \pm 0.11$
& $40.63 \pm 0.08$
& $40.20 \pm 0.20$
& $39.51 \pm 0.27$ \\[2pt]
J1549 & SDSS
& $40.25 \pm 0.04$
& $40.00 \pm 0.03$
& $39.47 \pm 0.14$
& $40.62 \pm 0.01$
& $40.23 \pm 0.01$
& $40.01 \pm 0.03$ \\
& DESI
& $39.98 \pm 0.25$
& $39.47 \pm 0.06$
& $39.82 \pm 0.07$
& $40.28 \pm 0.01$
& $39.83 \pm 0.02$
& $39.69 \pm 0.03$ \\[2pt]
J1620 & SDSS
& $39.81 \pm 0.07$
& $39.69 \pm 0.02$
& $39.22 \pm 0.16$
& $40.46 \pm 0.01$
& $40.16 \pm 0.01$
& $40.02 \pm 0.01$ \\
& DESI
& $39.78 \pm 0.19$
& $39.84 \pm 0.02$
& $39.83 \pm 0.03$
& $40.48 \pm 0.00$
& $40.11 \pm 0.01$
& $39.99 \pm 0.01$ \\[2pt]
J2203 & SDSS
& $40.05 \pm 0.20$
& $39.95 \pm 0.37$
& $39.98 \pm 0.26$
& $40.53 \pm 0.05$
& $40.44 \pm 0.05$
& $40.29 \pm 0.12$ \\
& DESI
& $40.02 \pm 0.30$
& $40.06 \pm 0.07$
& $40.52 \pm 0.07$
& $40.75 \pm 0.03$
& $40.27 \pm 0.08$
& $39.93 \pm 0.12$
\enddata
\tablecomments{
Narrow-line luminosities are reported as
$\log_{10}(L/\mathrm{erg\,s^{-1}})$.
[\ion{O}{2}] denotes the summed
$\lambda\lambda3726,3729$ luminosity,
[\ion{O}{3}] denotes $\lambda5007$,
[\ion{N}{2}] denotes $\lambda6584$, and
[\ion{S}{2}] denotes the summed
$\lambda\lambda6717,6731$ luminosity.
Quoted uncertainties are the formal $1\sigma$ measurement
uncertainties propagated from the spectral decomposition.
}
\end{deluxetable*}

\clearpage
\section{Cross-survey Flux Calibration and Aperture Effects}
\label{subsec:systematic_tests}
The SDSS and DESI spectra were obtained with different instruments,
fiber apertures, spectral resolutions, and spectrophotometric
calibrations. These differences must be distinguished from intrinsic
narrow-line variability before the selected [\ion{O}{3}] changes can
be interpreted physically. We therefore assessed the NIE candidates
against the ordinary two-epoch behavior of the complete
variability-selected parent sample, to which the same spectral-fitting
and line-measurement procedures were applied.

For the stacked spectral comparisons, the DESI spectra were first
convolved to the instrumental resolution of the corresponding SDSS
spectra. Both epochs were then shifted to the rest frame using the DESI
redshift and resampled onto a common wavelength grid. We extracted
three diagnostic windows containing
H$\beta$+[\ion{O}{3}],
[\ion{O}{2}], and
H$\alpha$+[\ion{N}{2}]+[\ion{S}{2}].
Within each window, a local continuum, $C_{\lambda}$, was fitted
independently, and the spectra were normalized as
\[
\widetilde{F}_{\lambda}
=
\frac{F_{\lambda}-C_{\lambda}}{C_{\lambda}}.
% \label{eq:continuum_normalization}
\]

This local normalization suppresses broad-band differences in
spectrophotometric calibration while preserving changes in the line
excess relative to the neighboring continuum. The same procedure was
applied to the NIE hosts and to the full parent population. We adopted
equal source weights for the fiducial stacks and repeated the
calculation with inverse-variance weighting as a robustness test.

Fiber aperture is a separate concern. The original SDSS spectra were
obtained through $3\arcsec$-diameter fibers, whereas DESI uses
$1\farcs5$-diameter fibers. Because low- and high-ionization emission
can have different radial distributions, the smaller DESI aperture may
alter both absolute line fluxes and line ratios. However, aperture loss
alone would generally reduce the later-epoch flux from spatially
extended emission. It cannot naturally explain a factor-of-several
increase in the absolute [\ion{O}{3}] flux measured through the
smaller DESI fiber. Nevertheless, a more compact
[\ion{O}{3}]-emitting component could be weighted differently from
more extended low-ionization gas, so the aperture effect must be tested
empirically rather than rejected from the fiber sizes alone.

We performed two complementary parent-sample tests. First, we examined
the evolution in the [\ion{N}{2}]-based BPT diagram, whose line ratios,
[\ion{O}{3}]/H$\beta$ and [\ion{N}{2}]/H$\alpha$, involve lines close
in wavelength and are therefore comparatively insensitive to
broad-band spectrophotometric errors. For this test, we selected parent
sources with median spectral signal-to-noise ratio
${\rm S/N}>10$ in both the H$\beta$ and H$\alpha$ fitting windows and
a narrow H$\beta$ detection at ${\rm S/N}>3$, yielding 151 objects.
This high-quality parent subset shows no coherent population-wide
migration toward higher excitation between the SDSS and DESI epochs
{(Figure~\ref{fig:supp_parent_bpt})}.

Second, we compared the fractional changes in [\ion{O}{2}] and
[\ion{O}{3}] for individual sources. After excluding the 19 NIE hosts,
we required ${\rm S/N}>5$ measurements of both oxygen lines in both
surveys, leaving 288 non-NIE parent sources. For each line, we defined
\[
\Delta_{\rm line}
=
\frac{
F_{\rm DESI}({\rm line})
-
F_{\rm SDSS}({\rm line})
}{
F_{\rm SDSS}({\rm line})
}.
% \label{eq:fractional_line_change}
\]

The non-NIE sources define the ordinary SDSS--DESI distribution
produced by measurement uncertainty, intrinsic variability, aperture
differences, and residual calibration effects. In this
[\ion{O}{2}]--[\ion{O}{3}] variability plane, the NIE hosts occupy an
extreme high-[\ion{O}{3}] tail while exhibiting substantially smaller
changes in [\ion{O}{2}]. They do not follow the locus expected from a
common multiplicative rescaling of the two lines
{(Figure~\ref{fig:ext_parent_phasespace})}.

The agreement among the parent-sample stacks, the nearby-line BPT
test, and the source-by-source oxygen-line comparison shows that the
NIE selection is not produced by a generic SDSS--DESI calibration
offset. The absolute increase in [\ion{O}{3}] through the smaller DESI
fiber also disfavors simple aperture loss as the dominant explanation.
Residual aperture-dependent effects may contribute to the scatter of
individual measurements, but they do not reproduce the coherent and
ionization-dependent response described in
Section~\ref{subsec:line_response}.

\renewcommand{\thefigure}{B\arabic{figure}}
\setcounter{figure}{0}

\begin{figure*}[t]
\centering
\includegraphics[width=\linewidth]
{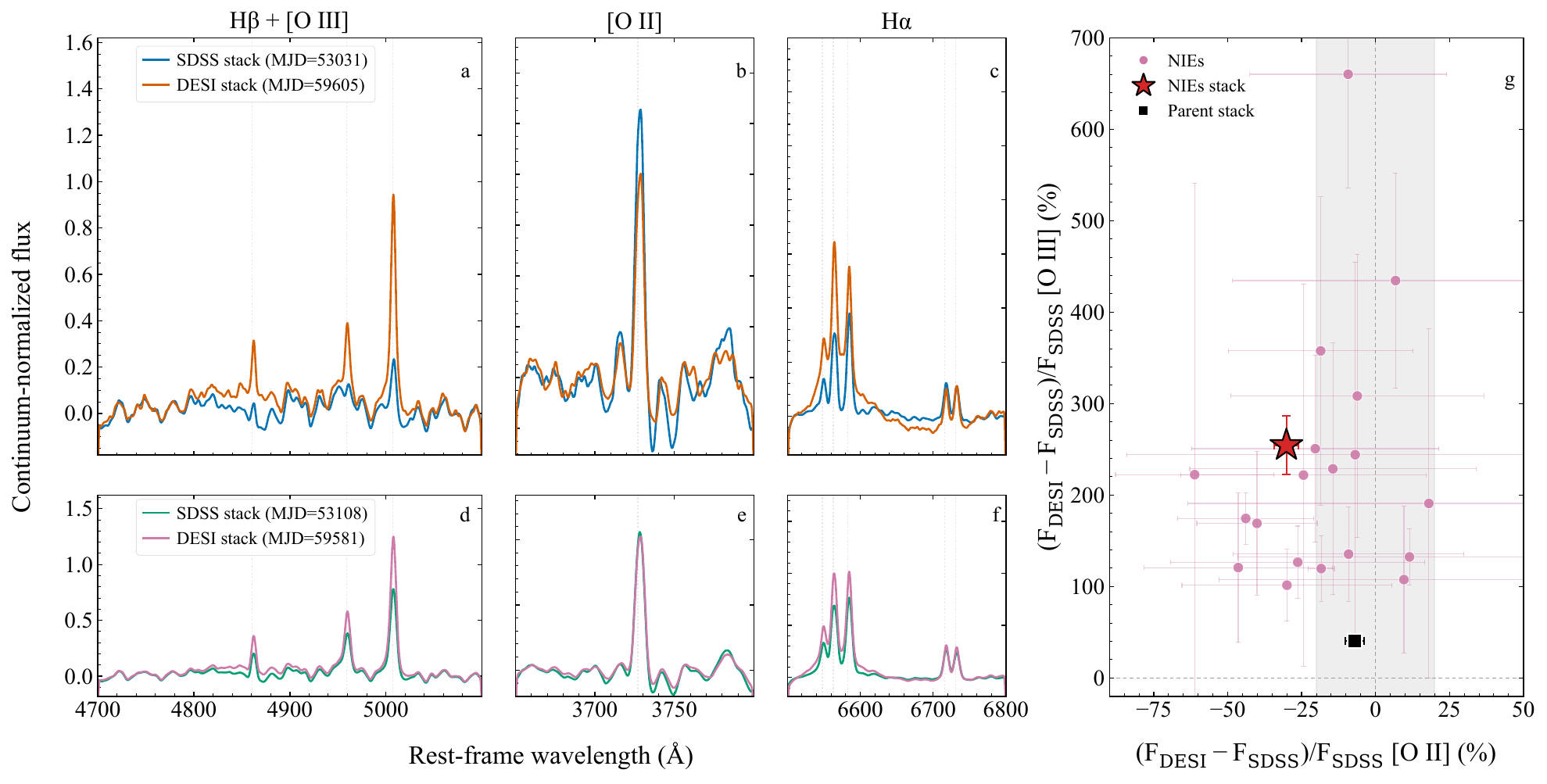}
\caption{
Robustness of the stacked spectral evolution to the stacking
estimator.
Inverse-variance-weighted stacked spectra of the 19 NIE hosts at the
SDSS (blue) and DESI (red) epochs.
(a)--(c) H$\beta$+[\ion{O}{3}], [\ion{O}{2}], and
H$\alpha$+[\ion{N}{2}]+[\ion{S}{2}] windows, corresponding to
Figure~\ref{fig:stacked_spectra}(a)--(c). The strong
[\ion{O}{3}] enhancement and weak [\ion{O}{2}] response reproduce the
equal-weight results, showing that the selective response is not
sensitive to the adopted stacking estimator.
}
\label{fig:ext_weighted_stack}
\end{figure*}

\begin{figure}[t]
\centering
\includegraphics[width=0.48\textwidth]
{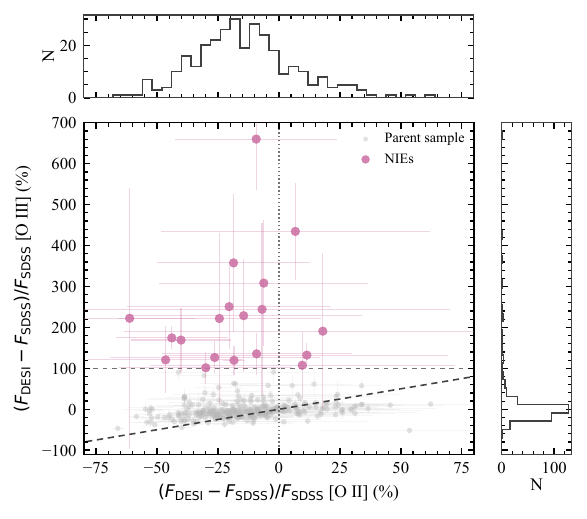}
\caption{
Parent-sample phase space of [\ion{O}{3}] and [\ion{O}{2}]
variability.
Gray points show the parent comparison sample of 288 non-NIE sources
with robust two-epoch measurements of both lines
(${\rm S/N}>5$ in each survey). Magenta points show the 19 NIE hosts.
The axes give the fractional line-flux changes,
$\Delta F/F_{\rm SDSS}
=(F_{\rm DESI}-F_{\rm SDSS})/F_{\rm SDSS}$,
for [\ion{O}{2}] and [\ion{O}{3}]; error bars represent $1\sigma$
uncertainties. Marginal histograms show the corresponding
distributions. The diagonal dashed line indicates equal fractional
changes in the two lines. The horizontal dashed line marks the
factor-of-two [\ion{O}{3}] selection threshold,
$\Delta F_{\rm [O\,III]}/F_{\rm SDSS}=1$, and the vertical dotted
line marks no [\ion{O}{2}] change. The NIE hosts occupy the
high-[\ion{O}{3}] tail while exhibiting comparatively modest
[\ion{O}{2}] evolution.
}
\label{fig:ext_parent_phasespace}
\end{figure}

\begin{figure}[t]
\centering
\includegraphics[width=0.48\textwidth]
{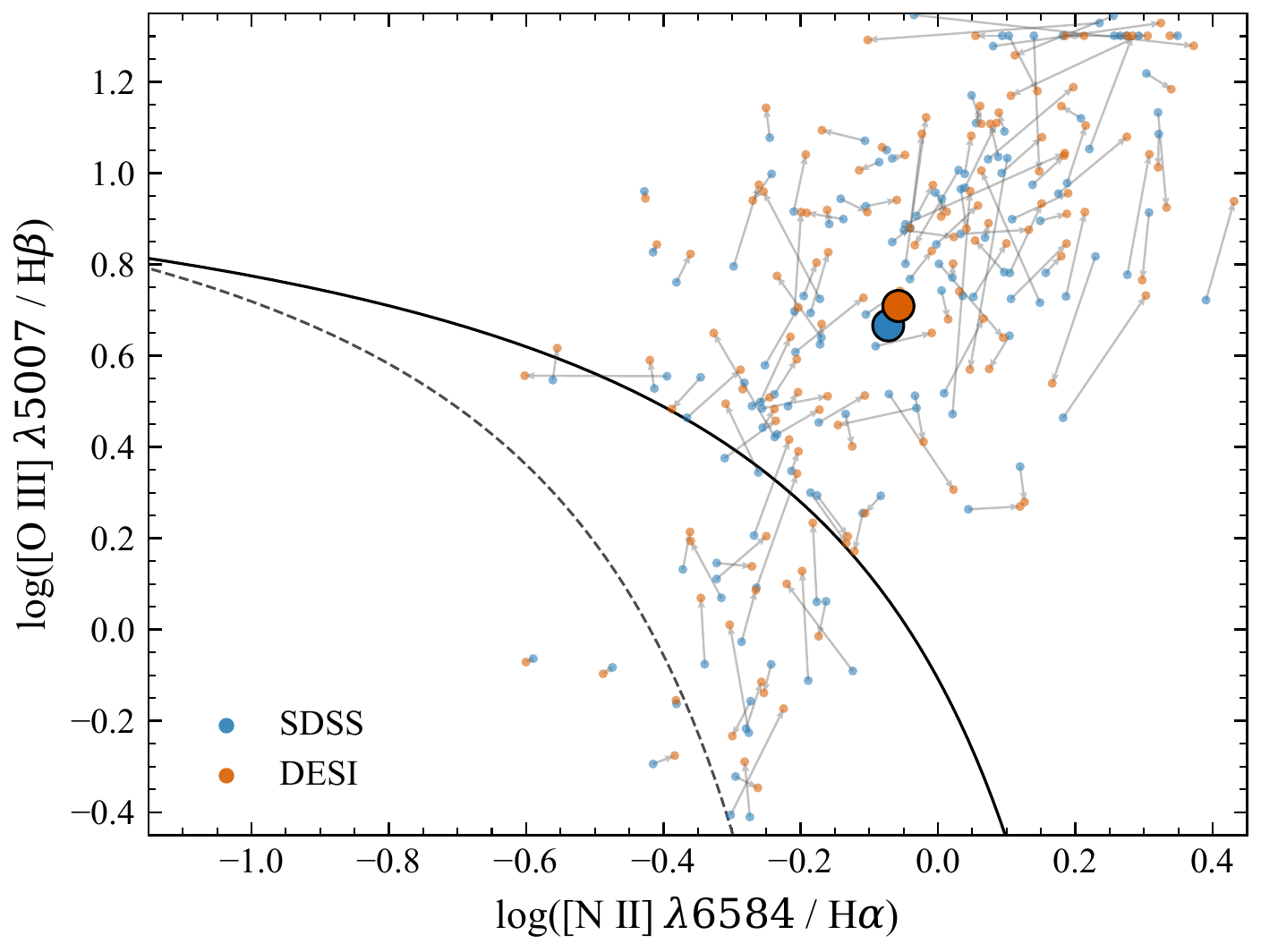}
\caption{
Parent-sample BPT diagram at the SDSS and DESI epochs.
The [\ion{N}{2}]-based BPT diagram is shown for the high-S/N parent
subset used to characterize ordinary two-epoch line-ratio scatter.
Sources are required to have median continuum
${\rm S/N}>10$ in both the H$\beta$ and H$\alpha$ fitting windows and
an H$\beta$ line detection at ${\rm S/N}>3$, leaving 151 sources.
Blue and orange points show the SDSS and DESI measurements,
respectively; gray segments connect the two epochs, and large symbols
mark the mean location of the sample at each epoch. The solid and
dashed curves show the theoretical and empirical maximum-starburst
boundaries of \citet{Kewley2001} and \citet{Kauffmann2003},
respectively. The parent sample exhibits no coherent migration toward
higher excitation.
}
\label{fig:supp_parent_bpt}
\end{figure}

\section{Ionizing-Photon Fractions across the AGN SED Range}
\label{app:agn_sed_range}

The representative AGN model with \(\alpha=-2.2\) reproduces the mean
DESI-epoch position, whereas harder continua extending to
\(\alpha=-1.4\) are required to encompass the full distribution of
DESI-epoch line ratios in the BPT plane
(Figure~\ref{fig:supp_multi_diagnostic_migration}). This range should
therefore be interpreted as the spread in ionizing spectral shapes
required for the population, rather than as a uniquely determined SED
for every source. Across the interval from \(\alpha=-2.2\) to
\(\alpha=-1.4\), the cumulative fraction of hydrogen-ionizing photons
above the O\(^{+}\rightarrow\)O\(^{2+}\) threshold increases from
\(\eta_{35}=12.4\%\) to \(26.4\%\), corresponding to a factor of
approximately 2.1. The fraction above the
He\(^{+}\rightarrow\)He\(^{2+}\) threshold increases from
\(\eta_{54}=4.7\%\) to \(14.2\%\), corresponding to a factor of
approximately 3.0. Thus, progressively harder AGN continua increase
the relative supply of photons above both ionization thresholds, with
a stronger enhancement in the extreme-ultraviolet tail. These harder
spectra provide the additional high-energy photons required to span
the most highly excited DESI-epoch sources beyond the
population-mean solution.

\renewcommand{\thefigure}{C\arabic{figure}}
\setcounter{figure}{0}
\begin{figure*}[t]
    \centering
    \includegraphics[width=\textwidth]{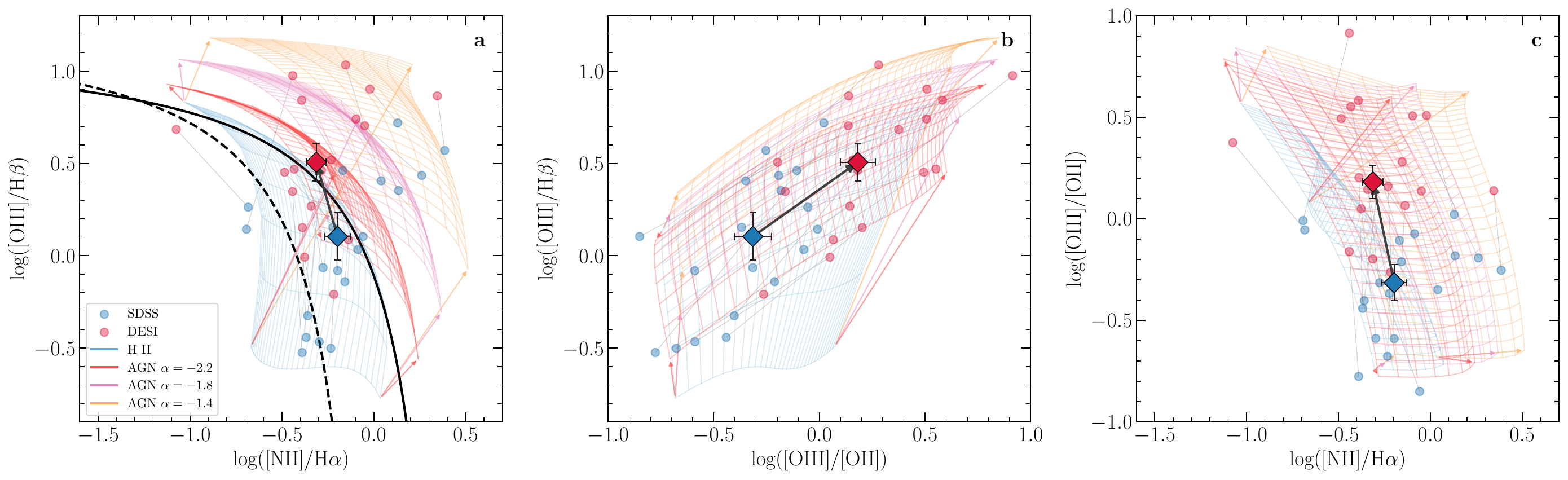}
    \caption{
    Multi-diagnostic narrow-line evolution between the SDSS and DESI epochs.
    Blue and red circles show the measurements of the 19 NIE hosts at the
    SDSS and DESI epochs, respectively, and light-gray segments connect the
    two epochs for each source.
    Large diamonds mark the sample-mean positions at the two epochs, and the
    black arrow indicates the mean displacement from SDSS to DESI.
    Panel (a) shows the classical [\ion{N}{2}]-based BPT diagram, panel (b)
    shows the O32--O3 plane, and panel (c) shows the O32--N2 plane.
    The colored grids show representative
    photoionization models.
    The blue grid is based on a stellar-population spectral energy
    distribution, while the orange, purple, and dark-red grids show AGN
    spectral energy distributions with power-law indices
    $\alpha=-2.2$, $-1.8$, and $-1.4$, respectively.
    The observed evolution is characterized by a coherent increase in O3 and
    O32 and a systematic migration across all three diagnostic planes.
    The range of AGN model grids required to encompass the DESI measurements
    indicates that changes in the ionizing spectral energy distribution, in
    addition to changes in $U_{\rm H}$, are needed to explain the narrow-line
    evolution.
    }
    \label{fig:supp_multi_diagnostic_migration}
\end{figure*}

% =====================================================================
% Table 3: MAPPINGS photoionization-grid parameters
% =====================================================================

\bibliography{sample701}{}
\bibliographystyle{aasjournalv7}

\end{document}